\documentclass[journal,10pt]{IEEEtran}
\ifCLASSINFOpdf
  \usepackage[pdftex]{graphicx}
  \usepackage{amsthm,amsmath,amssymb}
\usepackage{makecell}
\usepackage{mathrsfs}
  \usepackage{float}
  \usepackage{algorithm}
\usepackage{algpseudocode}
\usepackage{amsmath}
\usepackage{bm}
\usepackage{color}
\usepackage{comment}
\usepackage{booktabs}
 
\else
\fi
\usepackage{leftidx}
\usepackage{epstopdf}
\usepackage{amsmath}
\usepackage{subfigure}
\usepackage{makecell}
\usepackage{multirow}
\usepackage{soul, color, xcolor}
\usepackage{stfloats}
\usepackage{multirow}
\usepackage{cite}

\begin{document}

\title{LLM4SG: Adapting Large Language Model for Scatterer Generation via Synesthesia of Machines
}

\author{Zengrui Han,~\IEEEmembership{Graduate Student Member,~IEEE,} Lu Bai,~\IEEEmembership{Senior Member,~IEEE,} Ziwei Huang,~\IEEEmembership{Member,~IEEE,} and \\Xiang Cheng,~\IEEEmembership{Fellow,~IEEE} 

\thanks{Z. Han, Z. Huang, and X. Cheng are with the State Key Laboratory of Photonics and Communications, School of Electronics, Peking University, Beijing, 100871, P. R. China (email: zengruihan@stu.pku.edu.cn, ziweihuang@pku.edu.cn, xiangcheng@pku.edu.cn).}
\thanks{L. Bai is with the Joint SDU-NTU Centre for Artificial Intelligence Research (C-FAIR), Shandong University, Jinan, 250101, P. R. China (e-mail: lubai@sdu.edu.cn).}}


\maketitle

		\maketitle

\begin{abstract}
In this paper, a novel large language model (LLM)-based method for scatterer generation (LLM4SG) is proposed for sixth-generation (6G) artificial intelligence (AI)-native communications.
To provide a solid data foundation, we construct a new synthetic intelligent sensing-communication dataset for Synesthesia of Machines (SoM) in vehicle-to-vehicle (V2V) communications, named SynthSoM-V2V, covering multiple V2V scenarios with multiple frequency bands and multiple vehicular traffic densities (VTDs).
Leveraging the powerful cross-modal representation capabilities of LLMs, LLM4SG is designed to capture the general mapping relationship from light detection and ranging (LiDAR) point clouds to electromagnetic scatterers via SoM.
To address the inherent and significant differences across multi-modal data, synergistically optimized four-module architecture, i.e., preprocessor, embedding, backbone, and output modules, are designed by considering sensing characteristics and electromagnetic propagation.
The embedding module achieves effective cross-domain alignment of the sensing-communication domain and the natural language domain.
The backbone network is adapted in a task-guided manner with low rank adaptation (LoRA), where a carefully selected subset of layers is fine tuned to preserve general knowledge and reduce training cost.
The proposed LLM4SG is evaluated for scatterer generation by benchmarking against ray-tracing (RT) and conventional deep learning models.
Simulation results demonstrate that the proposed LLM4SG achieves superior performance in both full-sample and cross-condition generalization testing. 
It significantly outperforms conventional deep learning models across different frequency bands, scenarios, and VTDs, and demonstrates the capability to provide the massive and high-quality channel small-scale fading data required by AI-native 6G systems.

\end{abstract}

\IEEEpeerreviewmaketitle

\section{Introduction}

\IEEEPARstart{A}{s}
wireless communication systems evolve toward higher frequencies and more dynamic scenarios, a deep understanding of the channel and accurate channel modeling become fundamental prerequisites for system design and algorithm development~\cite{cheng2022channel}. 
Among various channel characteristics, small-scale fading caused by multipath propagation reflects the interaction between radio waves and objects in the propagation environment~\cite{scatterering}. 
This phenomenon has a significant impact on instantaneous SNR, error probability, and channel capacity, and therefore plays a vital role in the design of many communication algorithms~\cite{Precoding,pdma,Blockage1,Blockage2,sage}.

In conventional communication system design, channel small-scale fading modeling has primarily served to support physical-layer algorithm development and provide a unified validation platform.
Extremely high accuracy was not required, as long as the main characteristics of channel small-scale fading could be captured.
Two main approaches have been adopted.
Statistical channel models generate scatterers from empirical distributions and simulate multipath propagation through virtual links, thereby producing data with moderate accuracy that is sufficient for algorithm testing~\cite{yang2019cluster,huangbuchong2}.
Ray-tracing (RT), as a deterministic method, achieves higher site-specific accuracy and physical consistency, but requires detailed three-dimensional (3D) maps and incurs considerable computational complexity~\cite{qd1,qd2}.
Overall, these approaches have been adequate for conventional system design and testing.
With the advent of artificial intelligence (AI)-native sixth-generation (6G) communication systems, however, the performance ceiling is fundamentally determined by the quality and scale of training data~\cite{tnse,LMMs}. 
This creates an urgent demand for massive, high-fidelity channel small-scale fading data. 
Statistical models, limited to reproducing macroscopic statistics, cannot deliver the required precision, while RT, though accurate, cannot generate data at scale due to its computational burden. 
Consequently, neither approach can satisfy the stringent requirements of AI-native 6G systems for large-scale, high-quality channel small-scale fading data.

To address the demand for massive and high-quality channel data, recent studies have explored the use of AI techniques, including machine learning and deep learning~\cite{AImodeling}.
Direct end-to-end generation of complete multipath parameters often suffers from limited stability and a lack of physical interpretability. A more robust approach is to first produce intermediate representations that are closely tied to multipath formation, such as clusters or scatterers, and then derive the corresponding channel data through subsequent processing~\cite{part}.
Huang et al.\cite{cluster2} generated power-angle-spectrum (PAS)-based clusters from measurement data. They identified clusters as energy concentrations in the spectrum domain using a data-driven spectral clustering method with heuristic thresholding, which enables tracking of cluster births, deaths, and motions.
To capture temporal coherence, the authors in~\cite{cluster3} redefined clusters via similarity of multipath component (MPC) trajectories and generated trajectory-consistent clusters. MPCs were tracked using a Kuhn-Munkres assignment algorithm and subsequently clustered via a kernel power density (KPD)-based method.
Addressing robustness under rapid 3D dynamics, the authors in~\cite{cluster4} designed a deep learning-based MPC tracking network (MPCTNet) and applied an LSTM-based model to generate geometry-aware cluster tracks.
However, cluster boundaries are determined by specific algorithms and thresholds, leading to inconsistent definitions across scenarios and relatively coarse granularity. In contrast, scatterers are physical entities that model the interaction between radio waves and objects, and thus determine the multipath parameters with finer granularity and greater physical consistency across scenarios and frequency bands. Generating such scatterer-level data from radio-frequency (RF) measurements alone, however, requires stringent delay and angular resolution, placing heavy demands on system bandwidth and array size~\cite{localization}. Moreover, the lack of explicit representation of the physical environment further increases complexity and limits accuracy. As a result, RF-only AI-based generation remains constrained by hardware overhead and measurement costs, and cannot efficiently produce the massive, high-quality multipath data required for AI-native 6G communication systems.


Intelligent agents in 6G systems will be equipped with multi-modal sensing devices, e.g., RGB-D camera and light detection and ranging (LiDAR), providing rich and fine-grained information about their surrounding environments. 
Building on this trend, inspired by synesthesia of human, Synesthesia of Machines (SoM)~\cite{som} was proposed to enable the intelligent integration of multi-modal sensing and communication.
Inspired by SoM, a cross-modal generation paradigm has been proposed utilizing the easily accessible sensing data to generate channel data.
To explore the mapping relationship from physical environment to the electromagnetic space, the authors in~\cite{xgd} utilizes multi-modal data collected by cameras and LiDAR to generate the end-to-end path loss distribution.
In~\cite{TWC-Han}, we preliminarily explored the mapping relationship from LiDAR point clouds to scatterers via SoM.
The aforementioned studies are able to generate accurate data under specific scenarios but rely on conventional deep learning models with few parameters, which constrains its generalization potential to meet the demands of 6G AI-native communication systems.

Large language models (LLMs), which have revolutionized the field of natural language processing (NLP), have been successfully applied across domains~\cite{acrossdomain} and are capable of performing few-shot~\cite{few-shot} or zero-shot learning.
The semantic understanding enables the data generation in various fields, such as images~\cite{image}, videos~\cite{video}, tables~\cite{table}, and time-series data~\cite{time}.
LLMs exhibit significant adaptability on cross-modal tasks, such as channel prediction~\cite{LLM4CP,CSILLM}, beam prediction~\cite{beam}, CSI feedback~\cite{feedback}, and channel estimation~\cite{estimation} through fine-tuning, where smaller training datasets are adopted and fewer parameters are adjusted.
LLMs have demonstrated strong generalization and cross-modal representation abilities.
This paper focuses on a typical 6G scenario, namely vehicle-to-vehicle (V2V) communications, whose complex and highly dynamic nature poses significant challenges for cross-modal data generation.
Motivated by advances in LLMs, the first attempt is made to fine-tune an LLM for cross-modal generation from LiDAR point clouds to electromagnetic scatterers in V2V communications.
To be specific, built upon the architecture and design principles of SynthSoM dataset~\cite{synthsom}, we construct a new synthetic intelligent sensing-communication dataset for SoM in V2V communications (SynthSoM-V2V).
Based on SynthSoM-V2V dataset, a novel LLM-based method for Scatterer Generation (LLM4SG) from LiDAR point clouds is developed.
Leveraging the powerful capabilities of LLMs, the explored mapping relationship from LiDAR point clouds to scatterers exhibits robust generalizability across multiple V2V scenarios with multiple frequency bands and multiple vehicular traffic densities (VTDs). 
This capability positions it to effectively meet the evolving data demands of AI-native 6G communication systems.
It demonstrates the advantages of LLMs over conventional deep learning models in handling complex tasks and adapting to dynamic environments.
The major contributions and novelties of this paper are summarized as follows.
\begin{enumerate}
    \item To address the demand for massive and generalizable data in 6G AI-native communication systems, we explore cross-modal channel data generation from physical environment information. 
    Specifically, LLM4SG is proposed to adapt pre-trained LLMs to map LiDAR point clouds to electromagnetic scatterers for the first time.
    \item We develop SynthSoM-V2V, a comprehensive V2V dataset for intelligent sensing-communication integration. SynthSoM-V2V encompasses diverse V2V scenarios (e.g., urban and suburban) to meet various application scenario requirements, incorporates multiple frequency bands (e.g., 28 GHz and sub-6 GHz) due to the significant and nonlinear variations in propagation characteristics across frequencies, and includes multiple VTDs (e.g., low and high) as a critical factor influencing V2V channel properties. Unlike datasets designed for conventional deep learning models that focus on the specific scenario condition, the SynthSoM-V2V dataset is constructed to train and validate the accuracy and generalization capabilities of most of the related work across a wide range of scenario conditions.
    \item LLM4SG employs a synergistically optimized four-module architecture, i.e., preprocessor, embedding, backbone, and output modules, to adapt the LLM for the first time to explore complex nonlinear mapping relationship from LiDAR point clouds to scatterers. 
    Specifically, the embedding module is designed to achieve cross-domain alignment among sensing-communication domain and the natural language domain.
    A learnable frequency embedding is introduced for the first time, and scenario and traffic density information inherently contained in LiDAR point clouds is jointly embedded to enable generalization across frequencies, scenes, and VTDs.
    Building on this foundation, the backbone network is adapted with low rank adaptation (LoRA) in a task guided manner, freezing most parameters while fine-tuning a carefully selected subset.
    Together with the output module, this enables accurate scatterer generation across diverse scenarios, frequency bands, and VTD conditions.
    \item By virtue of precise alignment processing of multi-modal data and efficient adaptation to LLMs, LLM4SG delivers outstanding performance and generalization in scatterer generation. It achieves over 92\% accuracy in position generation and over 85\% in quantity generation. The innovative four-module architecture significantly enhances knowledge transfer across VTDs, frequency bands, and V2V scenarios, demonstrating superior generalization capabilities and outperforming conventional deep learning models by an average of over 7.5\%. LLM4SG is well-equipped to support the massive and diverse data requirements of AI-native communication systems.
\end{enumerate}

The remainder of this paper is organized as follows.
Section II introduces the synthetic intelligent sensing-communication dataset SynthSoM-V2V.
The LLM4SG, which can explore the generalized mapping relationship from LiDAR point clouds to scatterers, is developed in Section III.
In Section IV, simulation settings are illustrated and then the performance of the proposed LLM4SG method on full-sample and generalization testing is evaluated.
Finally, Section V draws the conclusion.

\section{SynthSoM-V2V Dataset Construction}

In this section, a new synthetic intelligent sensing-communication dataset for SoM in V2V communications, named SynthSoM-V2V, for multiple V2V scenarios with multiple frequency bands and multiple VTDs is constructed.
To achieve the synchronous collection of the sensing and communication data, two high-fidelity simulation software, i.e., Wireless InSite~\cite{InSite} and AirSim~\cite{shah2018airsim}, are intelligently incorporated as a integrated platform~\cite{CC,synthsom} to achieve in-depth integration and precise alignment.
Wireless InSite employs the RT technology to mimic radio wave propagation and collect channel data.
AirSim is an open-source platform developed on Unreal Engine and is used to collect high-fidelity sensing data, such as LiDAR point clouds. 
The SynthSoM-V2V dataset is designed to simulate diverse V2V communication conditions, which are challenging to capture in real-world measurements. 
It covers various scenarios, including urban and suburban, employs 5.9 GHz and 28 GHz frequency bands, and accounts for different traffic density levels, including low and high VTDs.
The SynthSoM-V2V dataset supports comprehensive testing and validation efforts, with the potential to contribute to broader research in V2V communications.
Specific scenario configurations and data collection are given as follows.

\subsection{Scenario Configuration}

Two typical V2V scenarios are constructed, i.e., an urban crossroad and a suburban forking road. 
The urban crossroad is surrounded by tall and densely packed buildings, which bring unique challenges, such as significant signal obstruction and multipath propagation. 
In contrast, buildings are fewer and more spread out in the suburban forking road, resulting in line-of-sight (LoS) communications and different propagation characteristics.
Urban and suburban scenarios encompass rich and sparse scattering environments, which provide a comprehensive representation of typical V2V communications.
This diversity not only captures the distinct challenges encountered in different environment, but also enhances the generalization capability of the proposed model. As a result, the following observation is applicable across a broad spectrum of real-world V2V communication scenarios.

Physical environments for the two scenarios are precisely rendered in AirSim. 
Then, 3D models in physical environments are properly imported from AirSim into Wireless InSite to achieve precise alignment.
As a consequence, electromagnetic space, which aligns with the physical environment, is established in Wireless InSite.
Fig.~\ref{t1} and Fig.~\ref{t2} presents urban crossroad and suburban forking road scenarios under different VTDs in AirSim and Wireless InSite.
In Fig.~\ref{t1}(a) and Fig.~\ref{t2}(a), the high-fidelity physical environment of the crossroad and forking road in AirSim is illustrated. 
Fig.~\ref{t1}(b) and Fig.~\ref{t2}(b) depict the corresponding electromagnetic space in Wireless InSite, where a precise alignment with scenarios in AirSim is implemented.
Note that investigation of the impact of VTDs on V2V communications is vital for the resilient communication system design, which can  support sixth-generation (6G) applications, such as target localization and autonomous vehicle tracking. 
High VTDs result in increased interactions among vehicles, which can affect signal quality and communication reliability.
In this case, crossroad and forking road scenarios contain high and low VTDs.
The numbers of vehicles on the crossroad under high and low VTDs are 20 and 12, respectively. The numbers of vehicles on the forking road under high and low VTDs are 15 and 8, respectively.

\subsection{Data Collection}

\begin{figure*}[t]
\centering
\includegraphics[width=1\textwidth]{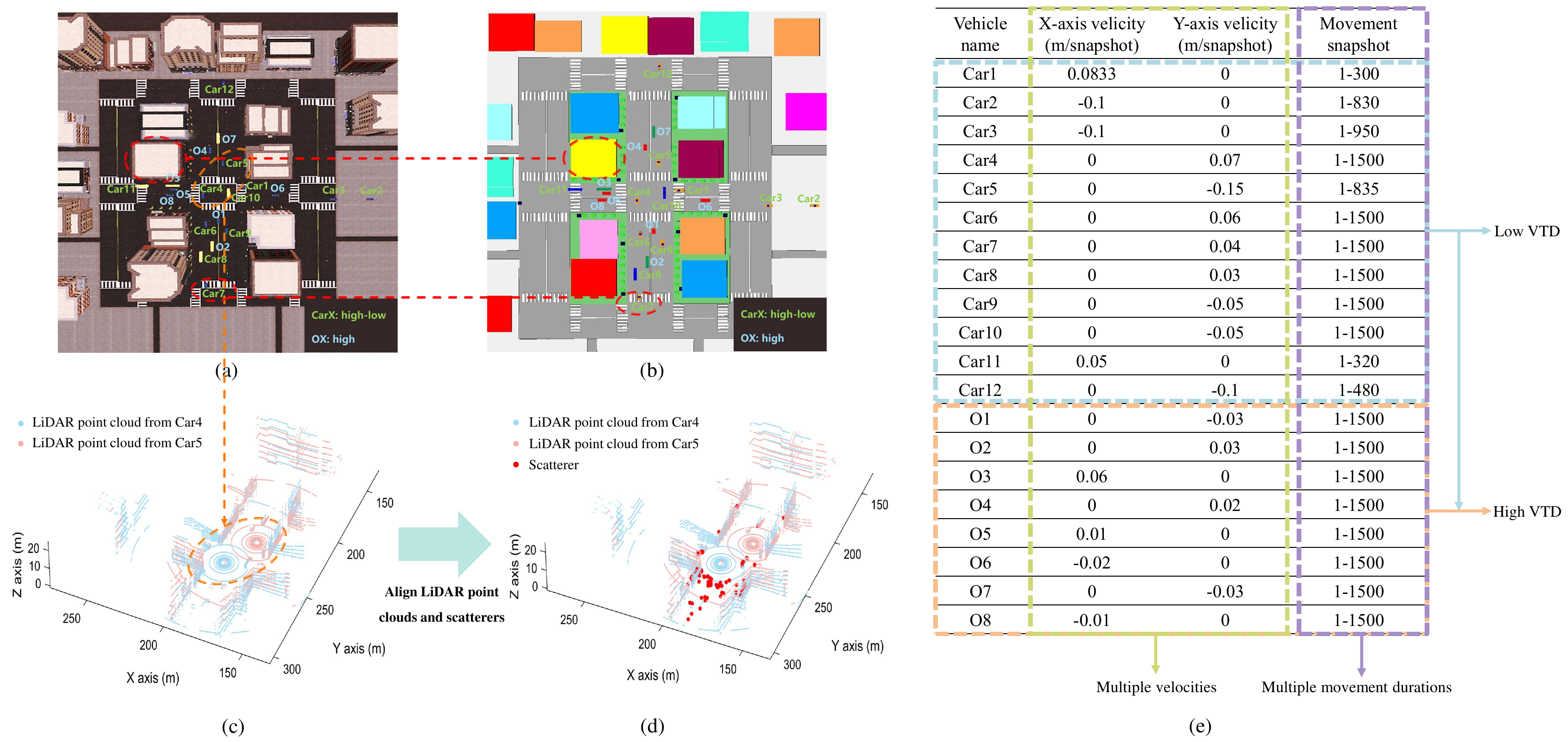}
\caption{Crossroads in AirSim and Wireless InSite and detailed parameter setting of vehicular trajectories under high and low VTDs. (a) Crossroad in AirSim under high and low VTDs at Snapshot 800. (b) Crossroad in Wireless InSite under high and low VTDs at Snapshot 800. (c) Combined LiDAR point clouds from Tx and Rx in AirSim. (d) Alignment between LiDAR point clouds and scatterers. (e) Detailed parameter setting of vehicular trajectories under high and low VTDs.}
\label{t1} 
\end{figure*}

\begin{figure*}[t]
\centering
\includegraphics[width=1\textwidth]{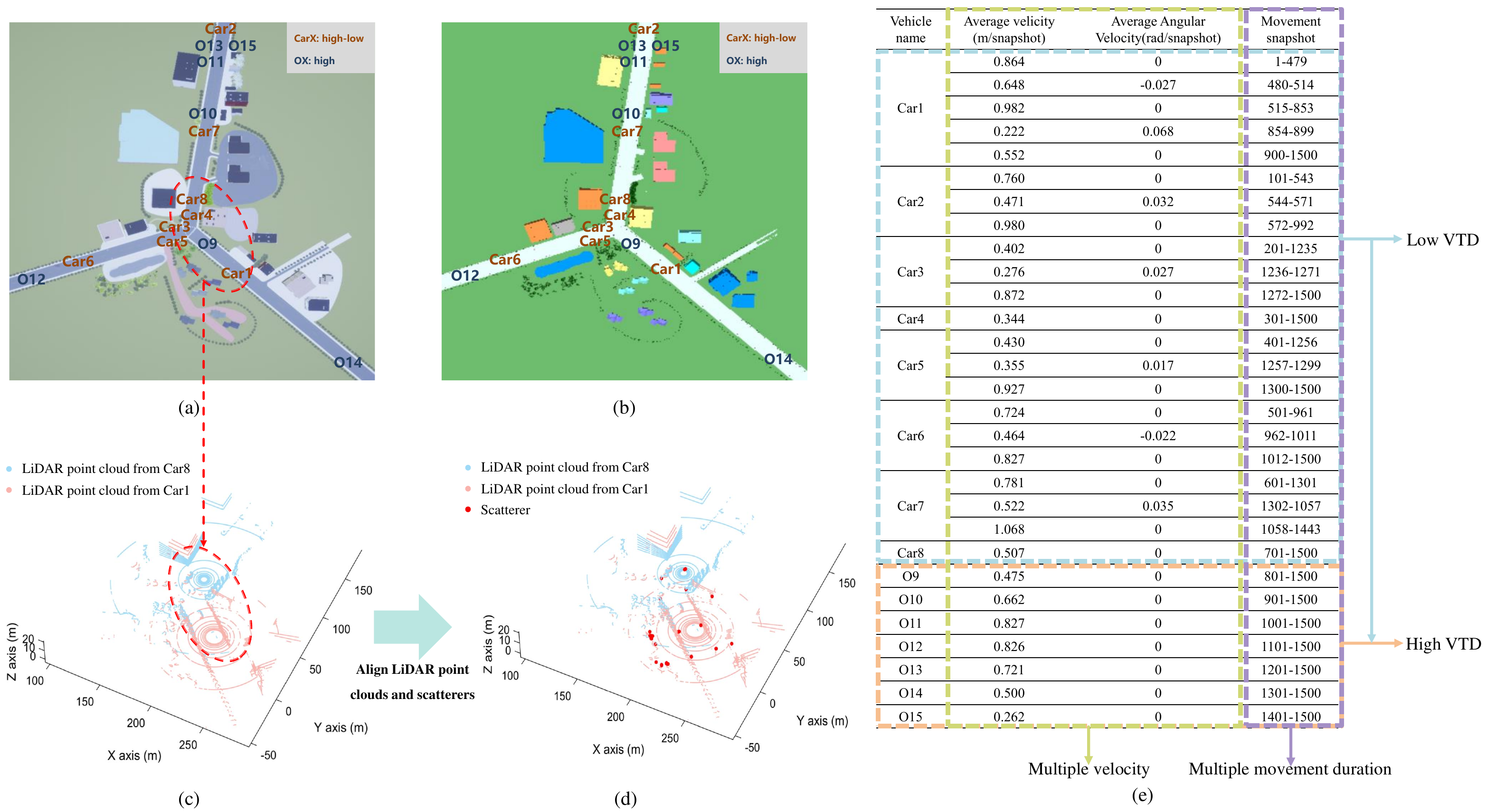}
\caption{Forking roads in AirSim and Wireless InSite and detailed parameter setting of vehicular trajectories under high and low VTDs. (a) Forking road in AirSim under high and low VTDs at Snapshot 1230. (b) Forking road in Wireless InSite under high and low VTDs at Snapshot 1230 (Vehicles which have not yet entered the scenario are labeled with initial positions). (c) Combined LiDAR point clouds from Tx and Rx in AirSim. (d) Alignment between LiDAR point clouds and scatterers. (e) Detailed parameter setting of vehicular trajectories under high and low VTDs.}
\label{t2}
\end{figure*}

The number of snapshots is set to 1500 for each VTD condition in each scenario.
First, initial positions of all vehicles are generated in AirSim, and the setup in Wireless InSite is precisely aligned with them.
Then, 3D coordinates of vehicles are set snapshot by snapshot in AirSim to mimic vehicular movement.
In Wireless InSite, by analyzing internal files and batch revising 3D coordinates of vehicles, vehicular trajectories are aligned snapshot by snapshot with those in AirSim.
For clarity, vehicular trajectories under high and low VTDs in the crossroad and forking road are shown in Fig.~\ref{t1}(c)~and Fig.~\ref{t2}(c).

Each vehicle follows a distinct trajectory, speed, and movement duration to construct dynamic and complex V2V scenarios. 
Additionally, each vehicle is equipped with a LiDAR and a communication unit, where sensor and transceiver positions and vehicular trajectories are precisely aligned. 
The LiDAR in AirSim features 16~channels with a scanning frequency of 10~Hz. 
High-fidelity LiDAR point clouds are obtained for each vehicle at each snapshot in AirSim.
The communication unit in Wireless InSite is operated with typical mmWave and sub-6~GHz frequency bands, i.e., $f_{\rm c}$ = 28~GHz carrier frequency with 2~GHz communication bandwidth and $f_{\rm c}$ = 5.9~GHz carrier frequency with 20~MHz communication bandwidth.
Each communication unit includes a transmitter (Tx) and a receiver (Rx), and the numbers of antennas at Tx and Rx are $M_{\rm T}$ = $M_{\rm R}$ = 1. 
By utilizing the RT technology, electromagnetic propagation scatterers and channel impulse response (CIR) data are generated for 18 pairs of V2V links under each VTD condition in the crossroad and forking road.

\section{LLM-Based Method for Scatterer Generation From LiDAR Point Clouds}

In this section, we propose LLM4SG, which is designed to generate scatterers from LiDAR point clouds leveraging the powerful cross-modal representation capabilities of LLMs, in order to meet the massive and diverse data requirements of AI-native communication systems.
LLM4SG, which contains preprocessor, embedding, backbone, and output modules, is a novel framework to achieve cross-domain alignment among sensing-communication domain and the natural language domain.
The preprocessor module refines the raw input LiDAR point clouds, preparing it for subsequent stages. 
In embedding module, a learnable frequency embedding is introduced for the first time, and scenario and VTD information inherently contained in LiDAR point clouds is jointly embedded to enable generalization across frequencies, scenarios, and VTDs.
Furthermore, the backbone, fine-tuned via LoRA in a task guided manner, extracts high-level features essential for accurate scatterer generation. 
These representations are subsequently transformed by the output module into precise scatterer grid maps.
The network components, as shown in Fig. \ref{network2}, and the training process are illustrated below.

\begin{figure*}[t]
\centering
\includegraphics[width=1\textwidth]{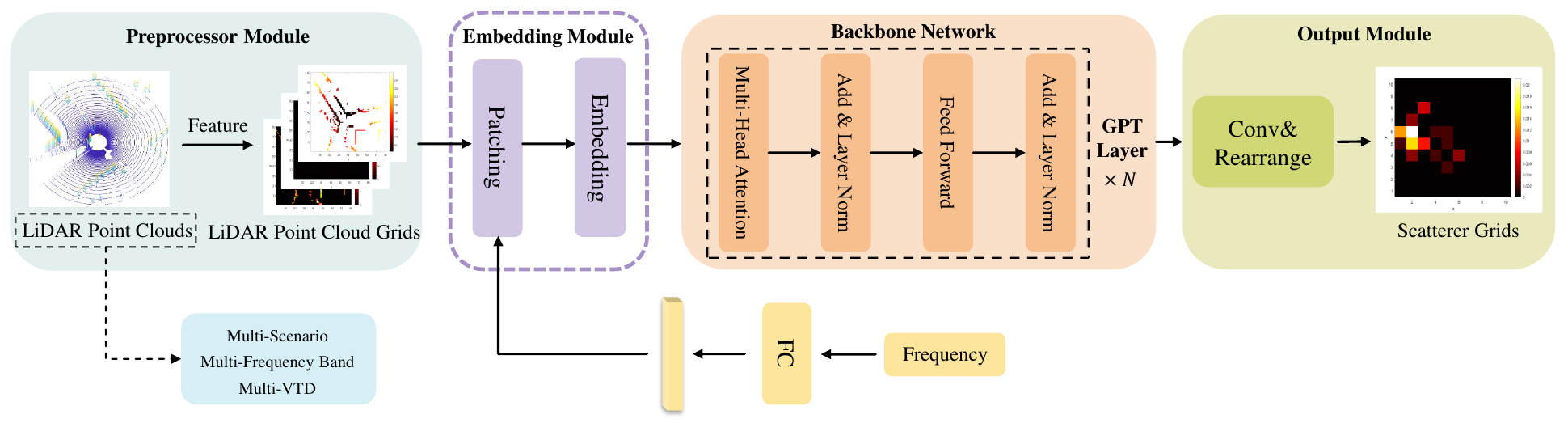}
\caption{An illustration of the network architecture of LLM4SG.}
\label{network2}
\end{figure*}

\subsection{Preprocessor Module}

The extensive data in LiDAR point clouds leads to high computational complexity, and noise and redundant points interfere with feature extraction and mapping relationship exploration. 
Meanwhile, there are inherent and significant differences between LiDAR point clouds and scatterers in terms of collection frequency bands and application significance resulting in the huge challenge of mapping relationship exploration.
To address the aforementioned challenges, a preprocessor module is applied to eliminate excess information and retain essential data.
The module includes concatenation and filtering of LiDAR point clouds and feature extraction.

In this process, to share the observed physical environment, LiDAR point clouds obtained from transceivers are concatenated.
Subsequently, ground points in LiDAR point clouds generated from the laser reflection off the ground are removed based on a threshold $H_{\rm G}$.
This step is essential as ground points typically contains a large portion of LiDAR point clouds and appear as a texture that significantly interferes with feature extraction.
Then, LiDAR point clouds undergo visibility region (VR)~\cite{huang2024lidar} filtering, which is widely used in channel modeling.
To be specific, scatterers are visible and contribute to channel characterization only if they are within the VR.
In~\cite{huang2024lidar}, the VR is defined as an ellipsoid with the Tx and Rx as its two focuses.
The major axis $2a(t)$ of the ellipsoid is calculated as the sum of the distances from the scatterers to transceivers.
The minor axis $2b(t)$ and the focal length $2c(t)$ are equal to the distance $D_{\rm T}(t)$ between transceivers.
As a result, processed LiDAR point clouds retain effective environmental information associated with vehicles, buildings, and trees within VR.
Processed LiDAR point clouds are expressed as
\begin{equation}
    \mathbb P(t)=\{\mathbf C(t)|\mathbf C(t)=[x_{\rm L}(t),y_{\rm L}(t),z_{\rm L}(t)]\}.
\end{equation}

Feature extraction of processed LiDAR point clouds is further conducted.
In fact, RT effectively identifies scatterers by simulating paths and interactions of rays in a 3D environment. 
Therefore, when extracting features from LiDAR point clouds, we consider factors similar to those in RT, such as object geometry, material properties, and interaction angles.
As a consequence, the widely used grid feature extraction technique~\cite{Jiang_2020_CVPR} in computer vision and image processing is leveraged.
LiDAR point clouds $\mathbb P(t)$ are divided into a grid map with a dimension of $m^{\rm x}\times m^{\rm y}$.
The length and width of each grid can be calculated as
\begin{equation}
    x_{\rm g}(t)=\frac{x_{\rm max}(t)-x_{\rm min}(t)}{m^{\rm x}}
\end{equation}
\begin{equation}
    y_{\rm g}(t)=\frac{y_{\rm max}(t)-y_{\rm min}(t)}{m^{\rm y}}
\end{equation}
where $x_{\rm min}(t)$, $x_{\rm max}(t)$, $y_{\rm min}(t)$, and $y_{\rm max}(t)$ are extreme values of LiDAR point clouds $\mathbb P(t)$ in $x$-axis and $y$-axis directions.
Each grid is determined by four coordinates, i.e., $(x_{\rm min}^{\rm g}(t),y_{\rm min}^{\rm g}(t))$, $(x_{\rm min}^{\rm g}(t),y_{\rm max}^{\rm g}(t))$, $(x_{\rm max}^{\rm g}(t),y_{\rm min}^{\rm g}(t))$, $(x_{\rm max}^{\rm g}(t),y_{\rm max}^{\rm g}(t))$.
LiDAR point clouds within each grid are given as 
\begin{equation}
\begin{aligned}
    \mathbb P^{\rm g}(t)=\{\mathbf C(t)|x_{\rm min}^{\rm g}(t) \leq x_{\rm L}(t)<x_{\rm max}^{\rm g}(t),\\
    y_{\rm min}^{\rm g}(t) \leq y_{\rm L}(t)<y_{\rm max}^{\rm g}(t)\}.
\end{aligned}
\end{equation}
Then, feature gird map $\mathbf F\in \mathbb R^{m^{\rm x}\times m^{\rm y}\times H}$ of LiDAR point clouds $\mathbb P(t)$ is extracted, where $H$ is the number of channels.
The first channel $\mathbf F^1$ of feature gird map $\mathbf F$ is the density of LiDAR point clouds in each grid, which can be calculated as
\begin{equation}
    \rho_{\rm g}(t)=\frac{|\mathbb P^{\rm g}(t)|}{x_{\rm g}(t)\times y_{\rm g}(t)}
\end{equation}
where $|\cdot|$ denotes the number of elements in a set.
The channel $\mathbf F^1$ represents a visual depiction of the scenario layout, which shows the distribution of scatterer-related elements, such as buildings, vehicles, and trees.
Next, the second channel $\mathbf F^2$ of feature gird map $\mathbf F$ is the maximum height of LiDAR point clouds in each grid, which can be calculated as
\begin{equation}
    h_{\rm g}(t)=\max_{\mathbf C(t)\in \mathbb P^{\rm g}(t)} z_{\rm L}(t).
\end{equation}
The channel $\mathbf F^2$ provides 3D information related to the environment.
Finally, the third channel $\mathbf F^3$ of feature gird map $\mathbf F$ is the distance from the transceiver to LiDAR point clouds in each grid, which can be calculated as
\begin{equation}
    d_{\rm g}(t)=\frac{d_{\rm g}^{\rm T}(t)+d_{\rm g}^{\rm R}(t)}{2}
\end{equation}
where $d_{\rm g}^{\rm T}(t)$ denotes the distance between the center of each grid and Tx and $d_{\rm g}^{\rm R}(t)$ denotes the distance between the center of each grid and Rx.
The channel $\mathbf F^3$ is similar to a position encoding layer, which further enhances the understanding of distance effects on scatterer distribution.

\subsection{Embedding Module}

In general, the mainstream LLM is based on the Transformer architecture~\cite{transformer}, which has demonstrated remarkable effectiveness in large-scale language tasks. 
The self-attention mechanism of Transformer enables the model to capture long-range dependencies in the text by attending to relevant tokens regardless of their position within the sequence.
In this architecture, the embedding module plays a key role in transforming LiDAR point cloud features and communication frequency information into a format optimized for the network. 
By projecting them into a fixed-dimensional latent space with positional encoding, the module ensures a consistent cross-modal representation across all layers and facilitates effective cross-domain alignment between the sensing-communication domain and the natural language domain.
To be specific, the embedding module includes patch embedding and positional embedding. 
The module further captures the global relationships between different regions of LiDAR point clouds while preserving positional information. 
This overcomes the receptive field limitations of conventional CNNs in image processing, enabling a more effective learning of electromagnetic propagation mechanisms.

The standard Transformer takes a one-dimensional (1D) sequence of token embeddings as input. 
\begin{figure*}[t]
\centering
\includegraphics[width=1\textwidth]{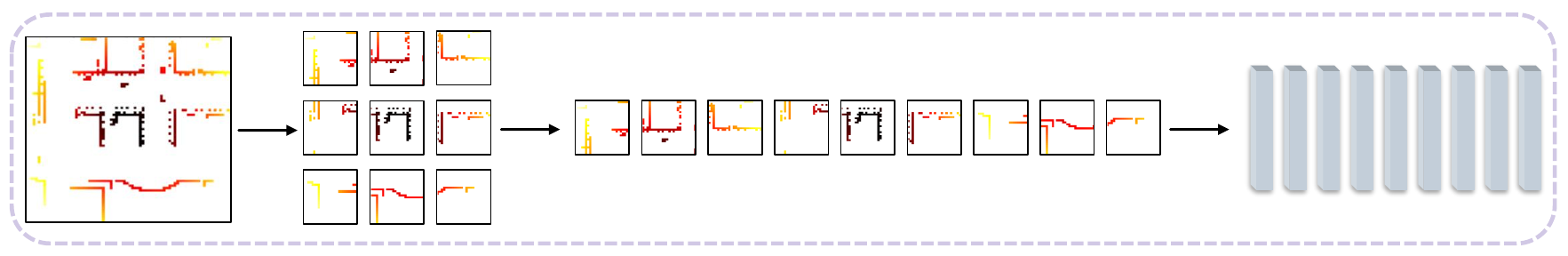}
\caption{An illustration of the patching operation of LiDAR point cloud features.}
\label{patch2}
\end{figure*}
For the two-dimensional (2D) LiDAR point cloud feature, we convert the feature grid map $\mathbf F\in \mathbb R^{m^{\rm x}\times m^{\rm y}\times H}$ into a sequence of flattened 2D patches $\mathbf F_{\rm p}$, as shown in Fig.~\ref{patch2}, which can be expressed as 
\begin{equation}
    \mathbf F_{\rm p}= \mathcal{V}(\mathcal{C}(\mathbf F)) \in \mathbb R^{N\times(Q^2\times H)}
\end{equation}
where $\mathcal{C}(\cdot)$ and $\mathcal{V}(\cdot)$ represent a convolutional layer and a reshape operation, respectively.
$(Q, Q)$ is the resolution of each patch and $N$ is the resulting number of patches, which can be calculated as 
\begin{equation}
    N=\frac{m^{\rm x}\times m^{\rm y}}{Q^2}.
\end{equation}
As the Transformer maintains constant latent vector size $D$ through all layers, patches are flattened and mapped to $D$ dimensions with a trainable linear mapping, which can be expressed as 
\begin{equation}
    \mathbf F_{\rm d}= \mathcal{L}(\mathbf F_{\rm p}) \in \mathbb R^{N\times D}
\end{equation}
where $\mathcal{L}(\cdot)$ represents a linear layer.

Since the Transformer processes inputs in parallel rather than sequentially, it lacks an inherent sense of token positions within a sequence.
By adding a position-based vector to each token, position embeddings allow the model to recognize relative/absolute positions of LiDAR point cloud tokens, which provide crucial context for understanding sequences.
\begin{figure}[t]
\centering
\includegraphics[width=2in]{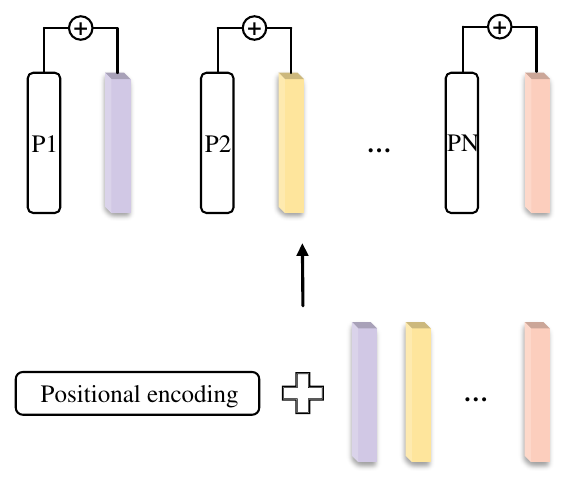}
\caption{An illustration of the position encoding.}
\label{positional2}
\end{figure}
A learnable 1D positional encoding $\mathbf{F}^{\rm PE}\in \mathbb R^{N\times 1}$ is implemented to facilitate the capturing of spatial information, which is shown in Fig.~\ref{positional2}. 
To incorporate positional encoding into patch embeddings, the positional encoding is broadcasted to the same dimension as $\mathbf{F}_{\rm d}$ and can be given by $\mathbf{\Tilde{F}}^{\rm PE}\in \mathbb R^{N\times D}$.
Then, positional encoding is added to the patch embeddings, which can be expressed as 
\begin{equation}
    \mathbf{F}^{\rm PE}_{\rm d}= \mathbf{F}_{\rm d} + \mathbf{\Tilde{F}}^{\rm PE}.
\end{equation}
Position embeddings enable the model not only to incorporate positional context but also to adaptively focus on the most relevant patches within the LiDAR point clouds.

Furthermore, due to huge differences in the collection frequency bands between sensing data and communication data, the communication frequency information is not inherently embedded in LiDAR point clouds.
However, the frequency information is crucial for accurate scatterer generation.
Consequently, the frequency information $f_{\rm c}$ is processed through a fully connected network and subsequently treated as a patch $\mathbf F_{\rm f}$, which is expressed as
\begin{equation}
    \mathbf F_{\rm f}=\mathcal{F}(f_{\rm c})\in \mathbb R^{1\times D}
\end{equation}
where $\mathcal{F}(\cdot)$ represents the fully connected network.
Then, the frequency patch is combined with all the other patches, which is expressed as 
\begin{equation}
    \mathbf{F}^{\rm PE}_{\rm d,f}=[\mathbf{F}^{\rm PE}_{\rm d},\mathbf F_{\rm f}]
    \in \mathbb R^{(N+1)\times D}.
\end{equation}

\subsection{Backbone Network}

Pre-trained LLMs are powerful tools capable of tackling various cross-modal downstream tasks by leveraging their vast knowledge base.
Compared to conventional deep learning models, LLMs offer superior ability to capture complex patterns, handle subtle cross-modal relationships, and deliver more robust/reliable performance.
For example, GPT-2~\cite{GPT} is built on the Transformer architecture, which leverages self-attention mechanisms to process and generate text. It consists of multiple stacked Transformer blocks, each comprising multi-head self-attention and feed-forward neural networks. This design enables GPT-2 to effectively capture long-range dependencies and contextual information.
The model takes tokenized text as input, embedding each token into a high-dimensional vector representation. These embeddings are passed through the stacked Transformer layers, where self-attention assigns context-dependent weights to tokens. Finally, a linear layer followed by a softmax function produces the probability distribution over the next token in the sequence.

Due to the inherent differences between sensing data and language data, GPT-2 cannot be directly applied to such tasks. 
To address this issue, analogous to text tokenization, sensing data are processed and embedded by the proposed preprocessor and embedding module to produce compatible representations.
The preprocessed LiDAR point cloud ``tokens" are then fed into the backbone of the GPT-2
\begin{equation}
    \mathbf{F}^{\rm LLM}=\mathcal{LLM}(\mathbf{F}^{\rm PE}_{\rm d,f})
\end{equation}
where $\mathcal{LLM}(\cdot)$ represents backbone networks, i.e., the first $N_{\rm L}$ layers of GPT-2.
In the proposed model, the GPT-2 layers can be replaced with other LLMs, as the LiDAR point clouds are transformed into a unified token format that ensures flexibility and compatibility. 
When selecting an alternative model, it is important to balance performance with computational overhead. 
This adaptability enables deployment across diverse AI-native communication system configurations and resource constraints.

\subsection{Output Module}

The output of the network is a single-layer grid map, which represents the spatial distribution of scatterers. 
The dimension of the grid map is $n^{\rm x}\times n^{\rm y}$ and the length and width of grid $g_{\rm s}^{x,y}(t)$ can be calculated as
\begin{equation}
    x_{\rm g}^{\prime}(t)=\frac{x_{\rm max}(t)-x_{\rm min}(t)}{n^{\rm x}}
\end{equation}
\begin{equation}
    y_{\rm g}^{\prime}(t)=\frac{y_{\rm max}(t)-y_{\rm min}(t)}{n^{\rm y}}.
\end{equation}
Scatterers within the VR are dropped onto the corresponding grid.
The density of scatterers in each grid is calculated as
\begin{equation}
    \phi_{\rm g}(t)=\frac{|\mathbb S^{\rm g}(t)|}{x_{\rm g}^{\prime}(t)\times y_{\rm g}^{\prime}(t)}
\end{equation}
where $\mathbb S^{\rm g}(t)$ is the scatterer set corresponding to each gird and $|\cdot|$ denotes the number of elements in a set.
The output module is designed to convert the output features of the LLM into the scatterer grid maps, which can be expressed as
\begin{equation}
    \hat{\boldsymbol{\phi_g}}(t)=\mathcal{C}(\mathcal{LR}(\mathcal{C}(\mathcal{V}(\mathbf{F}^{\rm LLM}))))
    \in \mathbb R^{n^{\rm x}\times n^{\rm y}}
\end{equation}
where $\mathcal{LR}(\cdot)$ represent a LeakyReLU operation~\cite{leakyrelu}.
Note that the cross-modal representation capabilities of LLMs hold promise for achieving higher-resolution generation of scatterer grids.

\section{Experiments}

In this section, the simulation settings are illustrated and then the performance of the proposed LLM4SG method is evaluated.

\subsection{Setup}

\subsubsection{Dataset Overview}
To organize the SynthSoM-V2V dataset, data entries will be named according to specific scenario conditions, including scenario name, frequency band, and VTD. Each dataset entry will reflect a unique combination of these conditions, which helps to clarify the selection of datasets in the following text. The naming convention follow a structure like ScenarioName\_FrequencyBand\_VTD, e.g., Crossroad\_28GHz\_HighVTD, ForkingRoad\_sub6GHz\_LowVTD.
The SynthSoM-V2V dataset for each condition includes 27,000 samples, comprising 18 transceiver pairs over 1500 snapshots.
In each sample, three-layer LiDAR point cloud feature grid maps are utilized as network input, while the single-layer scatterer grid map serves as the network output.

\subsubsection{Baseline}
The original model embedding module and backbone network are replaced with ResNet to establish a baseline for comparison. 
This modification utilizes ResNet’s deep residual connections, which facilitate effective feature capture and gradient flow. 
Using ResNet as the baseline provides a consistent reference point to evaluate performance improvements.

\subsubsection{Network and Training Configuration}
\begin{table}[t]
\centering
\renewcommand\arraystretch{1}
\caption{Hyper-Parameter for Network Design}
\label{parameter1}
\begin{tabular}{c|c}
\toprule 
\textbf{Parameter  }       & \textbf{Value} \\ \midrule
\begin{tabular}[c]{@{}c@{}}Size of LiDAR point cloud feature  \\ grid maps@$[3, m^{\rm x}, m^{\rm y}]$ \end{tabular}         &  [3, 80, 80]     \\ \midrule
Size of scatterer grid maps@$[1, n^{\rm x}, n^{\rm y}]$  &  [1, 10, 10]      \\ \midrule
\begin{tabular}[c]{@{}c@{}}Size of convolution kernels \\ in embedding module@$kernel\_size$\end{tabular} & (8, 8)      \\ \midrule
\begin{tabular}[c]{@{}c@{}}Size of convolution kernels \\ in output module@$kernel\_size$\end{tabular} &   (1, 1)     \\ \midrule
GPT-2 feature dimension@$D$  &   768 \\ \midrule
\begin{tabular}[c]{@{}c@{}}The number of GPT-2 layers@$N_{\rm L}$\end{tabular}  &  6 \\ \bottomrule 
\end{tabular}
\end{table}
Table \ref{parameter1} presents the hyper-parameters for the LLM4SG network, detailing the dimensions of the LiDAR point cloud feature grid maps, the dimensions of the scatterer grid map, and the architectural specifics of the ScaR network.
The smallest version of GPT-2, which has a feature dimension of F = 768, is employed, with only the first N = 6 layers being used.
\begin{table}[t]
\centering
\renewcommand\arraystretch{1}
\caption{Hyper-Parameter for Network Fine-Tuning}
\label{parameter2}
\begin{tabular}{c|c}
\toprule 
\textbf{Parameter  }       & \textbf{Value} \\ \midrule
Batch size      &  256     \\ \midrule
Starting learning rate  &  $1\times 10 ^{-2}$      \\ \midrule
Learning rate scheduler &   Per 150 epochs    \\ \midrule
Learning-rate decaying factor &   0.1    \\ \midrule
Epochs &   600  \\ \midrule
Optimizer & ADAM      \\ \midrule
Loss function & NMSELoss     \\ 
\bottomrule 
\end{tabular}
\end{table}
Table \ref{parameter2} outlines the hyper-parameters used for fine-tuning the LLM4SG network. The network is trained using PyTorch with the Adaptive Moment Estimation (ADAM) optimizer~\cite{diederik2014adam}.
The dataset for each condition is split into training, validation, and test sets in a 4:1:1 ratio. 
The model performance is assessed after training on the training set, and hyperparameters are tuned using the validation set. 
The test set evaluates the network's generalization ability on previously unseen data.
Note that in the pre-trained GPT-2 model, LoRA fine-tuning is applied only to specific target modules within the self-attention and feed-forward layers in a task guided manner. 
By introducing low-rank trainable adapters to these modules while keeping the original weights fixed, LoRA reduces the number of trainable parameters, substantially lowering training overhead and enabling efficient adaptation to the target task.
This adaptation enables the LLM to effectively transfer its pre-trained general knowledge to the scatterer generation task across diverse scenarios, VTDs, and frequency bands.
Meanwhile, it facilitates the rapid scatterer generation to meet massive and diverse data requirements of AI-native 6G communication systems.

\subsubsection{Loss Function}
In this model, the output is a $n^{\rm x}\times n^{\rm y}$ scatterer grid map. The loss function is based on the normalized mean squared error (NMSE) of each grid's predicted value compared to the ground truth, which can be calculated as
\begin{equation}
    NMSE(\boldsymbol{\phi_g},\hat{\boldsymbol{\phi_g}})=
    \frac{\sum_{i=1}^N (\phi_g-\hat{\phi_g})^2}{\sum_{i=1}^N \phi_g^2}.
\end{equation}
The NMSE provides a smooth and continuous gradient, which allows optimization algorithms to converge more efficiently during training.

\subsubsection{Performance Metric}

To evaluate the performance of LLM4SG, two aspects of performance need to be considered.
First, the evaluation focuses on the accuracy of predicting scatterer locations, namely, whether each cell in the grid map correctly indicates the presence of scatterers. 
The location evaluation metric can be expressed as 
\begin{equation}
    P_{\rm pos}=\frac{N_{\rm z}+N_{\rm nz}}{n^{\rm x}\times n^{\rm y}}
\end{equation}
where $N_{\rm z}$ represents the number of grids whose predicted values and ground truth are both zero and $N_{\rm nz}$ represents the number of grids whose predicted values and ground truth are both non-zero.
Second, the accuracy of predicting the number of scatterers within each grid is examined.
Specifically, the number evaluation metric is defined as the proportion of grids where the predicted number deviates from the ground truth by less than a specified threshold $S_{\rm th}$.
The number evaluation metric is calculated as
\begin{equation}
    P_{\rm num}=\frac{|\{g_{\rm s}^{x,y}(t)|
    \frac{|\phi_{\rm g}(t)-\hat{\phi_{\rm g}}(t)|}{\phi_{\rm g}(t)}<S_{\rm th}\}|}{|\{g_{\rm s}^{x,y}(t)\}|}
\end{equation}
where $|\cdot|$ denotes the number of elements in a set.


\subsection{Performance of LLM4SG Under Individual Conditions}
The performance of LLM4SG is evaluated under individual environmental conditions, including different scenarios, frequency bands, and VTDs. 
We performed experiments using various hyper-parameter values and selected the optimal results for comparison.
The $P_{\rm pos}$ and $P_{\rm num}$ results in the crossroad and forking road are shown in Table \ref{results1} and Table \ref{results2}, respectively.
\begin{table*}[t]
\renewcommand\arraystretch{1}
\setlength{\tabcolsep}{13.5pt}
\centering
\caption{Performance of the LLM4SG Under Various Environmental Conditions in Crossroad}
\label{results1}
\begin{tabular}{ccccccccccccc}
\hline
\multirow{2}{*}{Model} & \multicolumn{2}{c}{\begin{tabular}[c]{@{}c@{}}Crossroad\_28GHz\\ \_HighVTD\end{tabular}} & \multicolumn{2}{c}{\begin{tabular}[c]{@{}c@{}}Crossroad\_28GHz\\ \_LowVTD\end{tabular}} & \multicolumn{2}{c}{\begin{tabular}[c]{@{}c@{}}Crossroad\_sub6GHz\\ \_HighVTD\end{tabular}} & \multicolumn{2}{c}{\begin{tabular}[c]{@{}c@{}}Crossroad\_sub6GHz\\ \_LowVTD\end{tabular}} \\ \cline{2-9} 
& $P_{\rm pos}$          & $P_{\rm num}$                & $P_{\rm pos}$          & $P_{\rm num}$               & $P_{\rm pos}$           & $P_{\rm num}$              & $P_{\rm pos}$           & $P_{\rm num}$             \\ \hline
ResNet&91.5\%& 86.2\%&92.6\%&88.1\%& 88.5\%&80.2\% &91.8\% &85.2\% \\ \hline
LLM4SG   & \textbf{94.1\%}& \textbf{91.8\%}&\textbf{96.3\%}& \textbf{94.5\%}&\textbf{92.3\%} &\textbf{85.3\%}  & \textbf{97.1\%} & \textbf{94.0\%}\\ \hline
\end{tabular}
\end{table*}
\begin{table*}[t]
\renewcommand\arraystretch{1}
\setlength{\tabcolsep}{10pt}
\centering
\caption{Performance of the LLM4SG Under Various Environmental Conditions in Forking Road}
\label{results2}
\begin{tabular}{ccccccccccccc}
\hline
\multirow{2}{*}{Model} & \multicolumn{2}{c}{\begin{tabular}[c]{@{}c@{}}ForkingRoad\_28GHz\\ \_HighVTD\end{tabular}} & \multicolumn{2}{c}{\begin{tabular}[c]{@{}c@{}}ForkingRoad\_28GHz\\ \_LowVTD\end{tabular}} & \multicolumn{2}{c}{\begin{tabular}[c]{@{}c@{}}ForkingRoad\_sub6GHz\\ \_HighVTD\end{tabular}} & \multicolumn{2}{c}{\begin{tabular}[c]{@{}c@{}}ForkingRoad\_sub6GHz\\ \_LowVTD\end{tabular}} \\ \cline{2-9} 
& $P_{\rm pos}$          & $P_{\rm num}$                & $P_{\rm pos}$          & $P_{\rm num}$              & $P_{\rm pos}$           & $P_{\rm num}$               & $P_{\rm pos}$           & $P_{\rm num}$               \\ \hline
ResNet&88.1\%& 83.2\% & 89.9\% & 85.5\%&89.2\% &  84.7\% & 89.8\%& 85.8\% \\ \hline
LLM4SG   & \textbf{92.8\%}  & \textbf{90.8\%}&\textbf{94.1\%} & \textbf{91.7\%}&\textbf{92.7\%} & \textbf{90.3\%} &\textbf{95.9\%}&\textbf{91.9\%}  \\ \hline
\end{tabular}
\end{table*}
The charts reveal that LLM4SG maintains high performance across different environments, demonstrating strong robustness and adaptability.
Meanwhile, experimental results reveal that the performance varies under different environmental conditions. 
Specifically, the generation accuracy is slightly higher in suburban forking road compared to urban crossroad, due to the low signal interference and multipath effects in suburban. 
Moreover, the model achieves improved performance under low VTDs, attributed to the less congested environments and diminished interference, which simplify the channel characteristics compared to high VTD scenarios. 
Furthermore, the model performs better at lower frequencies, as high-frequency signals are more vulnerable to atmospheric absorption and propagation losses, making scatterer generation more difficult.

\subsection{Performance of LLM4SG in Scenario Condition Transfer}
To evaluate the generalization ability of LLM4SG, its performance across different frequency bands, scenarios, and VTD transfers is assessed. 
This generalization is essential to meet the massive and diverse data requirements of AI-native communication systems.
\subsubsection{Performance Comparison in Frequency Band Transfer}
\begin{figure}[t]
    \begin{minipage}[t]{0.5\linewidth}
        \centering
        \includegraphics[width=\textwidth]{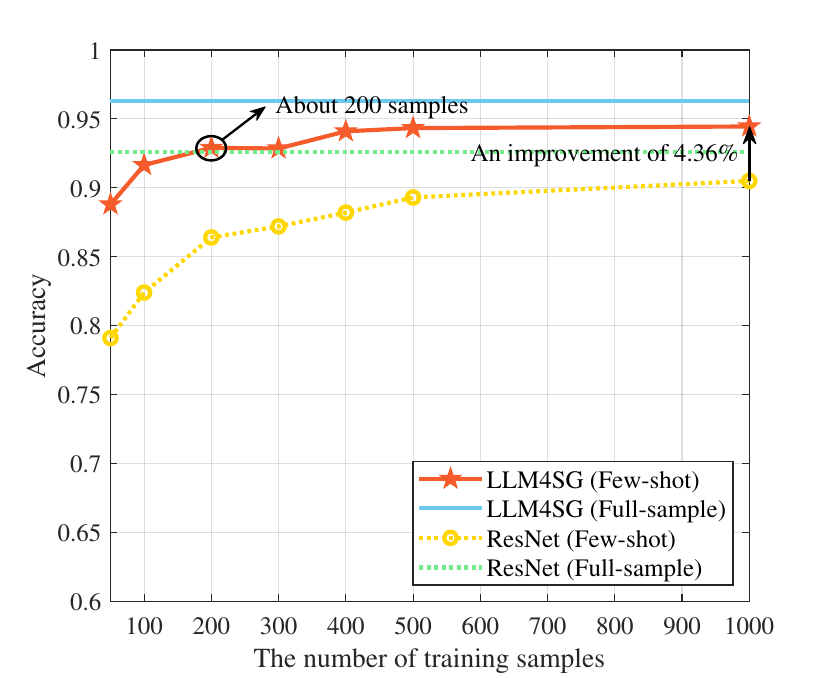}
        \centerline{(a)}
    \end{minipage}%
    \begin{minipage}[t]{0.5\linewidth}
        \centering
        \includegraphics[width=\textwidth]{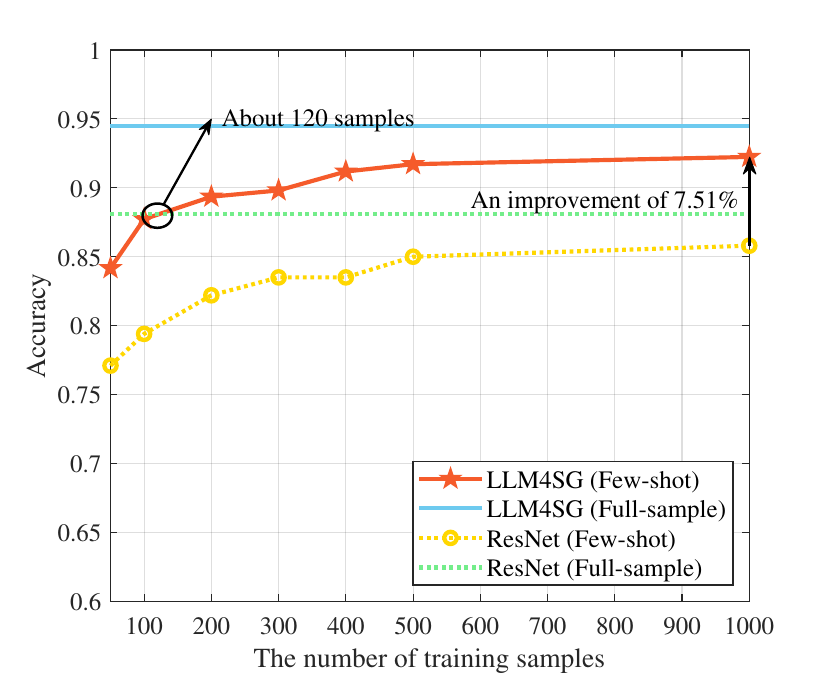}
        \centerline{(b)}
    \end{minipage}
    \hfill
    \begin{minipage}[t]{0.5\linewidth}
        \centering
        \includegraphics[width=\textwidth]{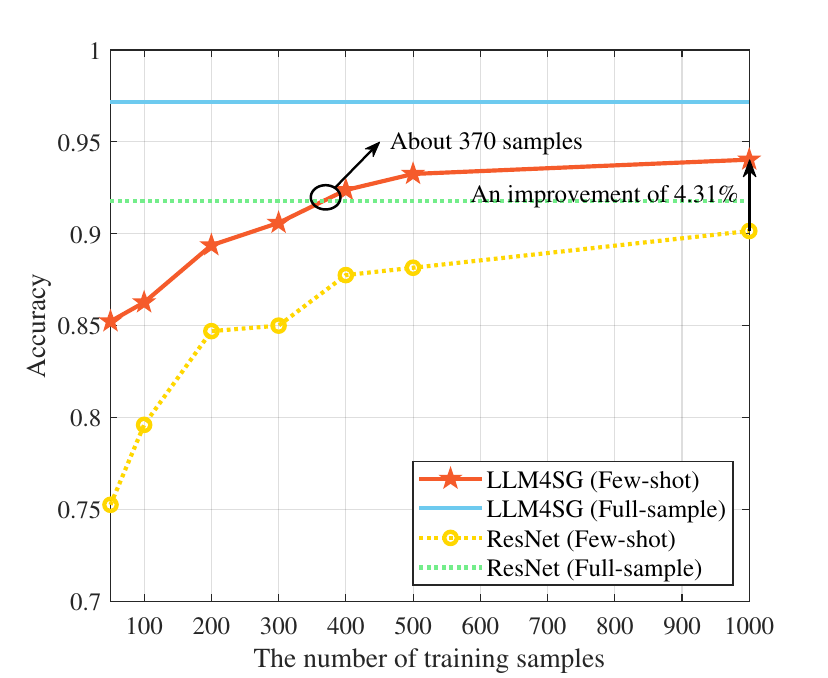}
        \centerline{(c)}
    \end{minipage}%
    \begin{minipage}[t]{0.5\linewidth}
        \centering
        \includegraphics[width=\textwidth]{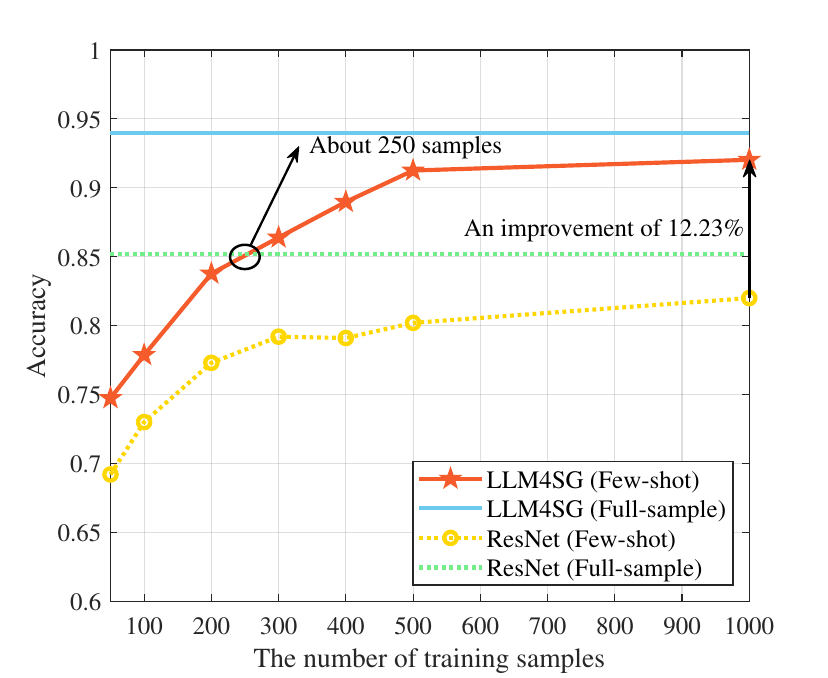}
        \centerline{(d)}
    \end{minipage}
    \caption{Generalization performance in frequency band transfer in urban crossroad. (a) Location evaluation metric from sub-6 GHz to 28 GHz. (b) Number evaluation metric from sub-6 GHz to 28 GHz. (c) Location evaluation metric from 28 GHz to sub-6 GHz. (d) Number evaluation metric from 28 GHz to sub-6 GHz.}
    \label{sub6to28}
\end{figure}

\begin{figure}[t]
    \begin{minipage}[t]{0.5\linewidth}
        \centering
        \includegraphics[width=\textwidth]{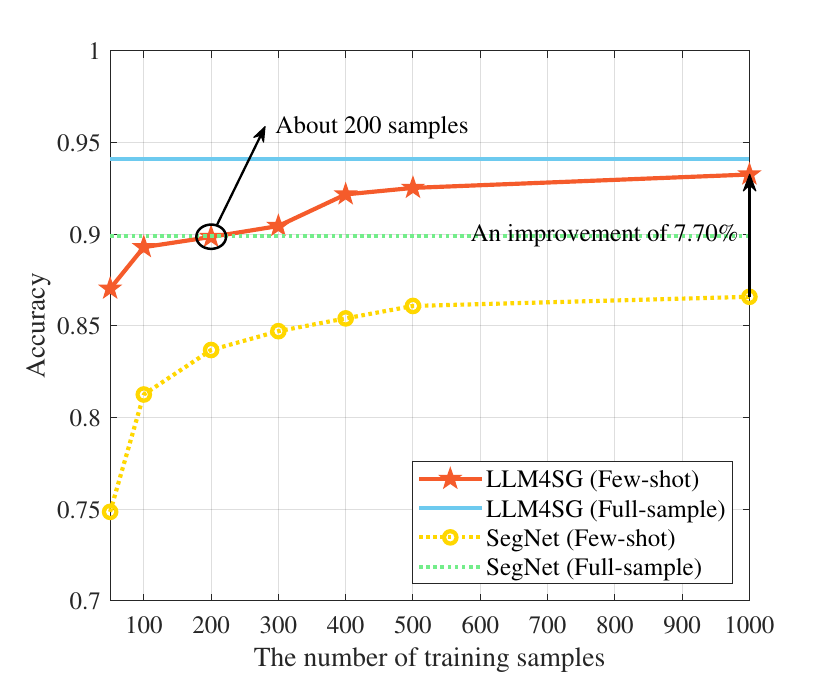}
        \centerline{(a)}
    \end{minipage}%
    \begin{minipage}[t]{0.5\linewidth}
        \centering
        \includegraphics[width=\textwidth]{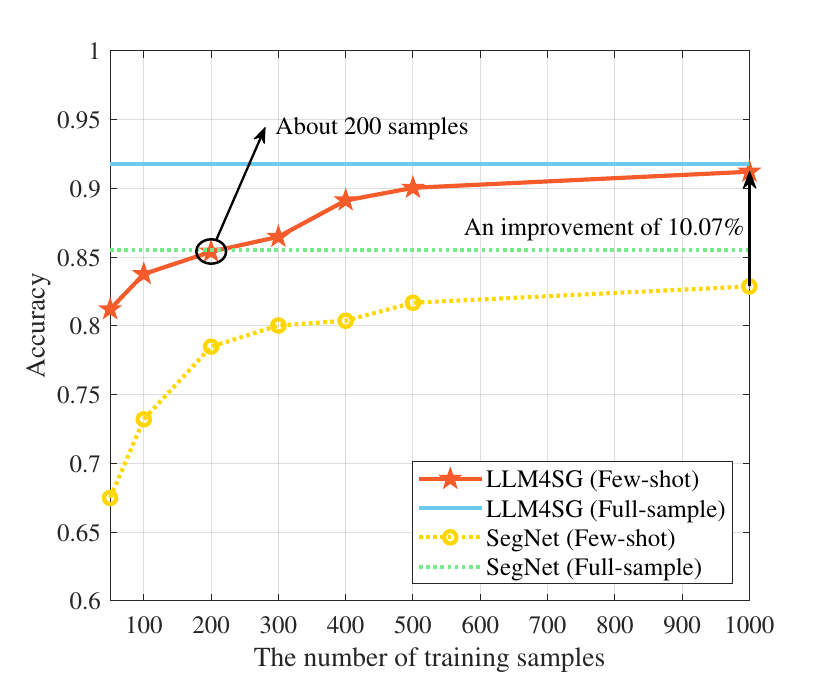}
        \centerline{(b)}
    \end{minipage}
    \hfill
    \begin{minipage}[t]{0.5\linewidth}
        \centering
        \includegraphics[width=\textwidth]{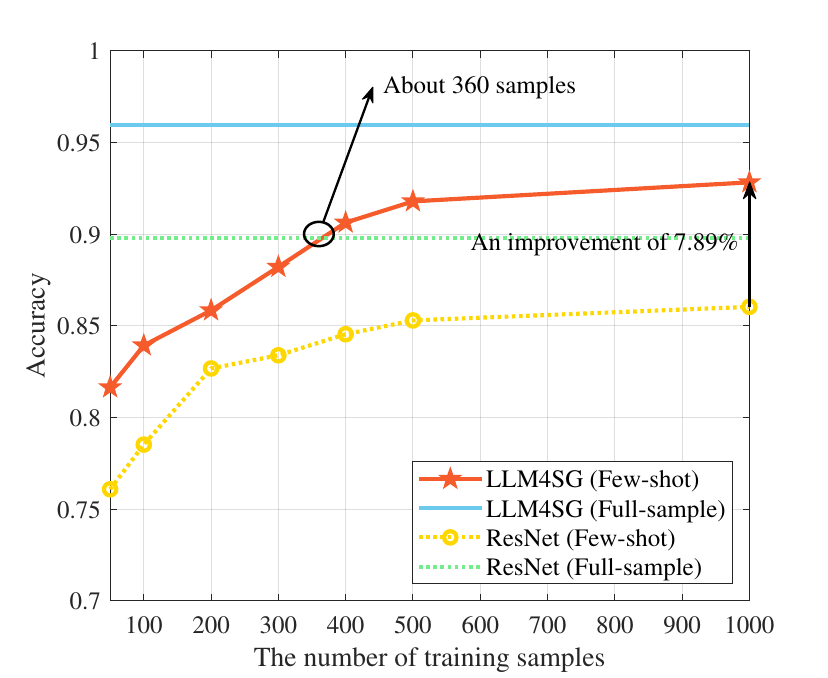}
        \centerline{(c)}
    \end{minipage}%
    \begin{minipage}[t]{0.5\linewidth}
        \centering
        \includegraphics[width=\textwidth]{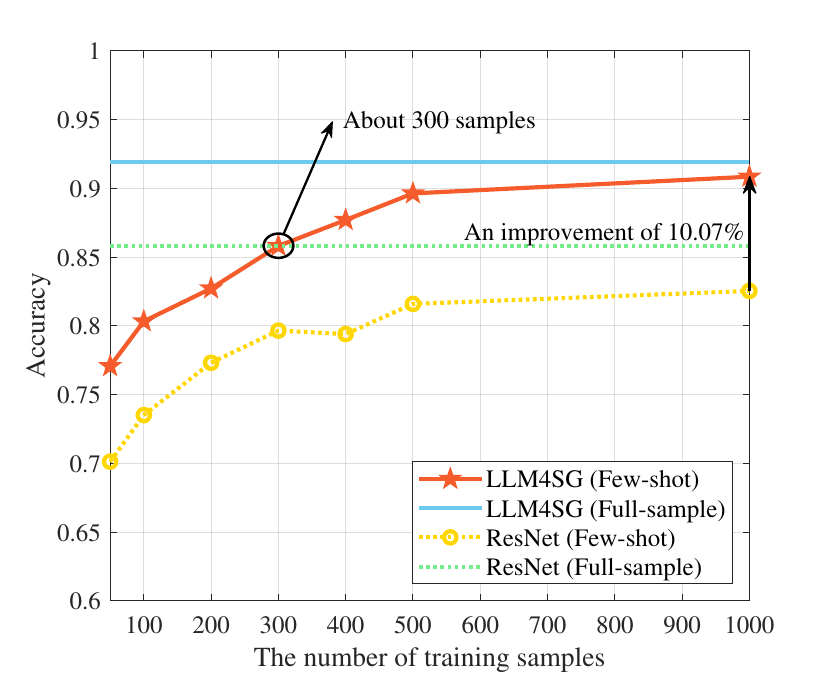}
        \centerline{(d)}
    \end{minipage}
    \caption{Generalization performance in frequency band transfer in suburban forking road. (a) Location evaluation metric from sub-6 GHz to 28 GHz. (b) Number evaluation metric from sub-6 GHz to 28 GHz. (c) Location evaluation metric from 28 GHz to sub-6 GHz. (d) Number evaluation metric from 28 GHz to sub-6 GHz.}
    \label{sub6to282}
\end{figure}

The performance of LLM4SG is evaluated in frequency band transfer learning between 28 GHz and sub-6 GHz bands. 
A few-shot learning approach is used, where the model is initially trained on one frequency band and then fine-tuned with a limited number of samples from the target band. 
This method assesses the adaptability of the model to distinct frequency bands with minimal additional data. 
The performance of LLM4SG is compared to that of the baseline across different few-shot sample numbers in crossroad and forking road, as shown in Fig. \ref{sub6to28} and Fig. \ref{sub6to282}.
Results indicate that, while both models benefit from few-shot adaptation, LLM4SG consistently outperforms the baseline, especially with minimal few-shot data. 
With fewer than 400 samples (less than 2\% of the full dataset) for few-shot training, LLM4SG can already match the baseline’s performance under full-sample training.
These results underscore the superior ability of LLM4SG to generalize effectively across the 28 GHz and sub-6 GHz bands with minimal tuning.
These results further show that transferring from sub-6 GHz to 28 GHz allows LLM4SG to surpass ResNet’s full-sample performance with far fewer samples, whereas the reverse transfer requires more data to achieve comparable results.
This asymmetry is due to the lower path loss and richer multipath characteristics of sub-6 GHz channels, which offer more informative and diverse training samples for cross-frequency generalization.
To sum up, this capability provides frequency diverse, high quality data for V2V scenarios and supports spectrum aware training and deployment in 6G AI native communication systems.

\subsubsection{Performance Comparison in Scenario Transfer}

\begin{figure}[t]
    \begin{minipage}[t]{0.5\linewidth}
        \centering
        \includegraphics[width=\textwidth]{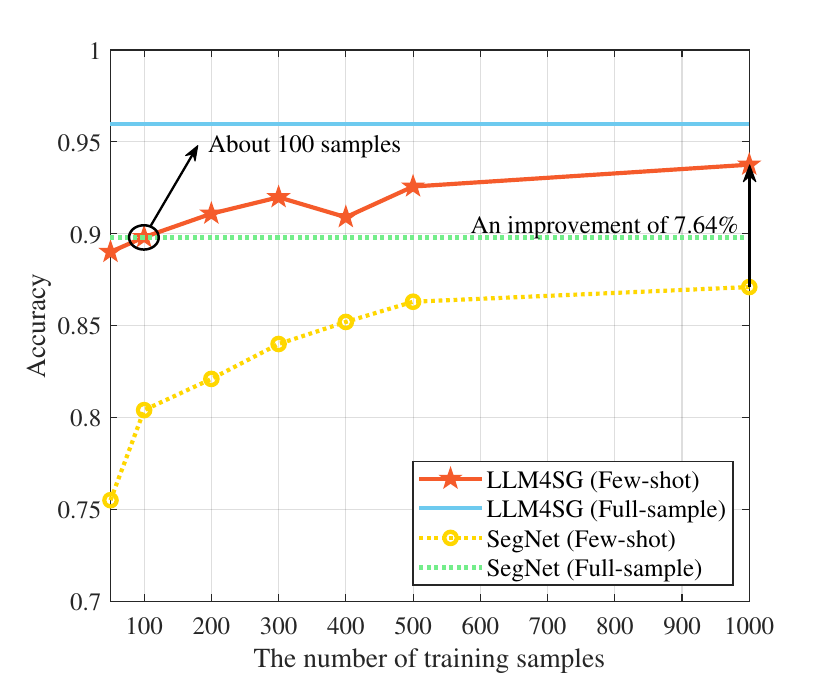}
        \centerline{(a)}
    \end{minipage}%
    \begin{minipage}[t]{0.5\linewidth}
        \centering
        \includegraphics[width=\textwidth]{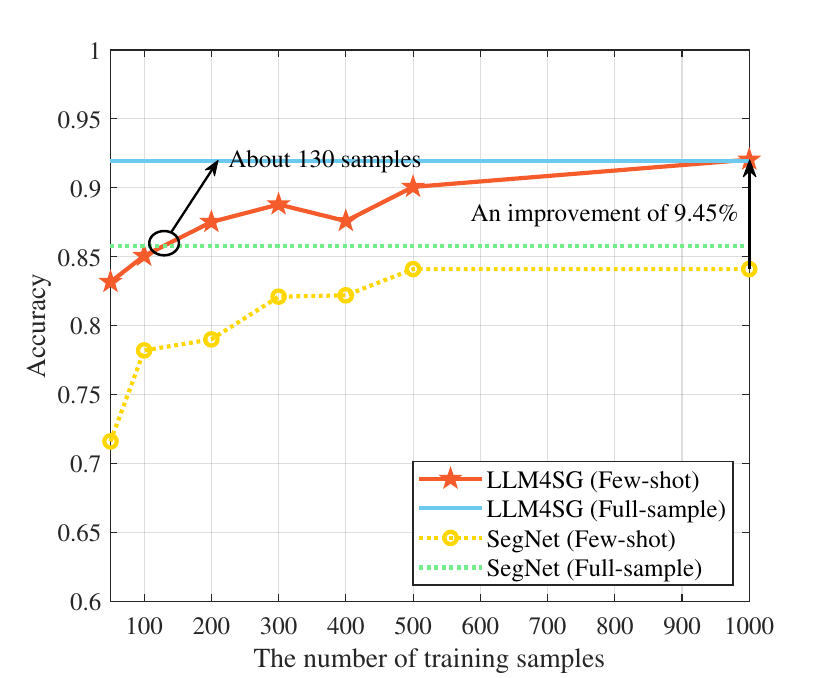}
        \centerline{(b)}
    \end{minipage}
    \hfill
    \begin{minipage}[t]{0.5\linewidth}
        \centering
        \includegraphics[width=\textwidth]{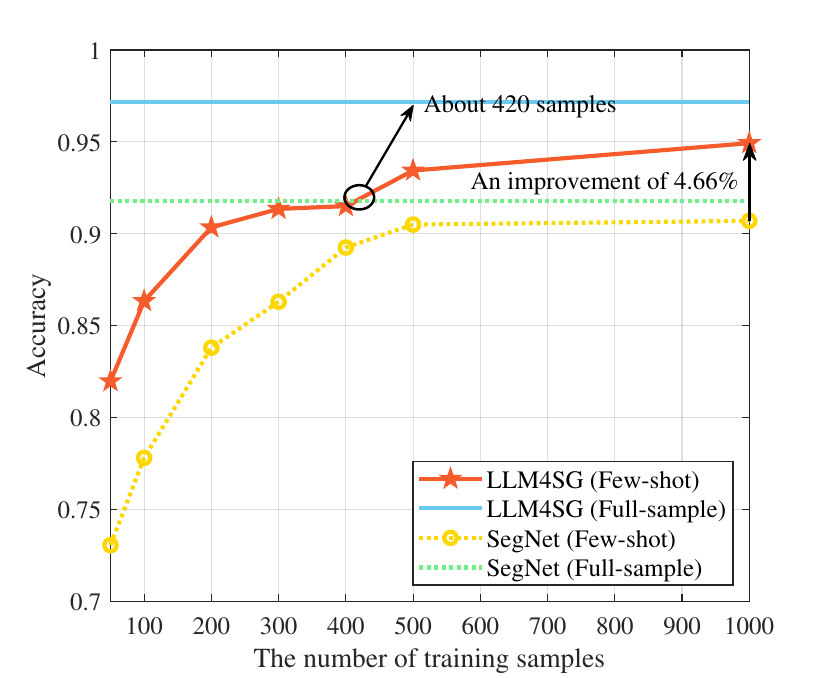}
        \centerline{(c)}
    \end{minipage}%
    \begin{minipage}[t]{0.5\linewidth}
        \centering
        \includegraphics[width=\textwidth]{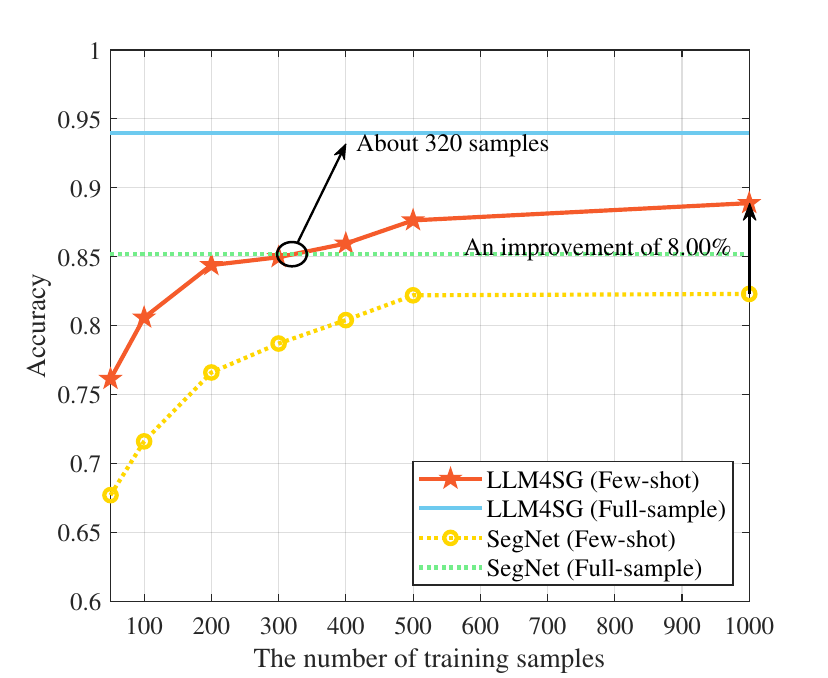}
        \centerline{(d)}
    \end{minipage}
    \caption{Generalization performance in scenario transfer. (a) Location evaluation metric from urban crossroad to suburban forking road. (b) Number evaluation metric from urban crossroad to suburban forking road. (c) Location evaluation metric from suburban forking road to urban crossroad. (d) Number evaluation metric from suburban forking road to urban crossroad.}
    \label{shi2fen}
\end{figure}

\begin{figure}[t]
    \begin{minipage}[t]{0.5\linewidth}
        \centering
        \includegraphics[width=\textwidth]{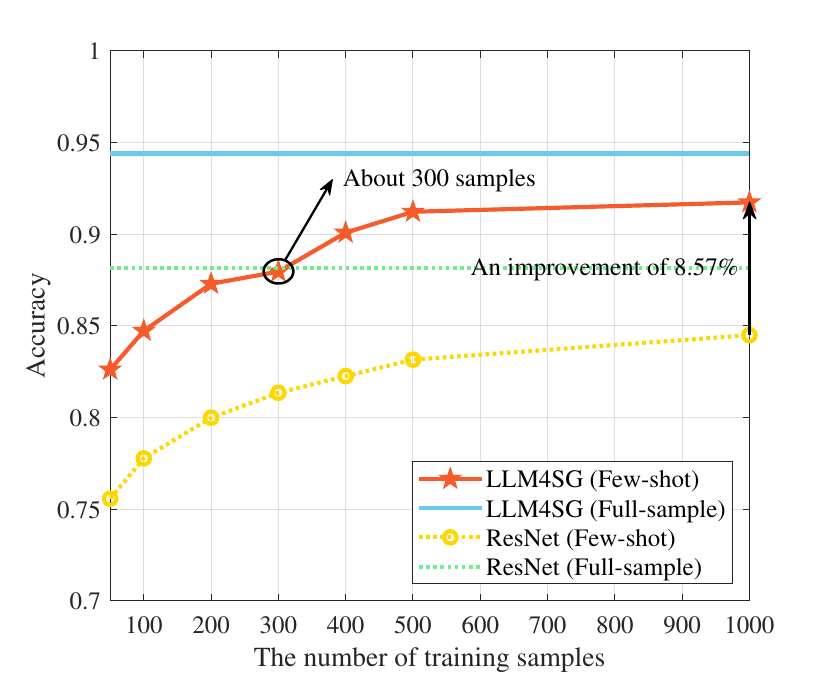}
        \centerline{(a)}
    \end{minipage}%
    \begin{minipage}[t]{0.5\linewidth}
        \centering
        \includegraphics[width=\textwidth]{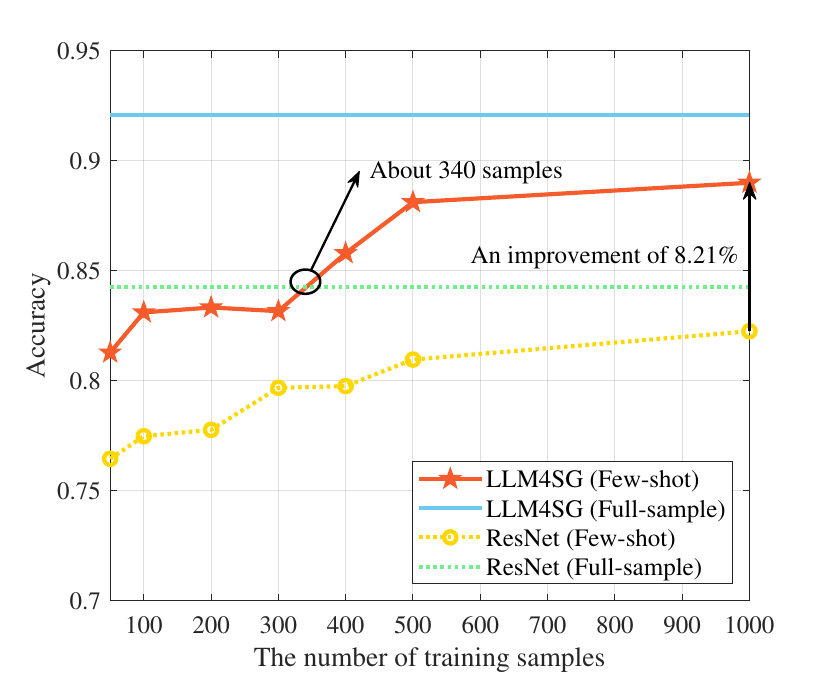}
        \centerline{(b)}
    \end{minipage}
    \hfill
    \begin{minipage}[t]{0.5\linewidth}
        \centering
        \includegraphics[width=\textwidth]{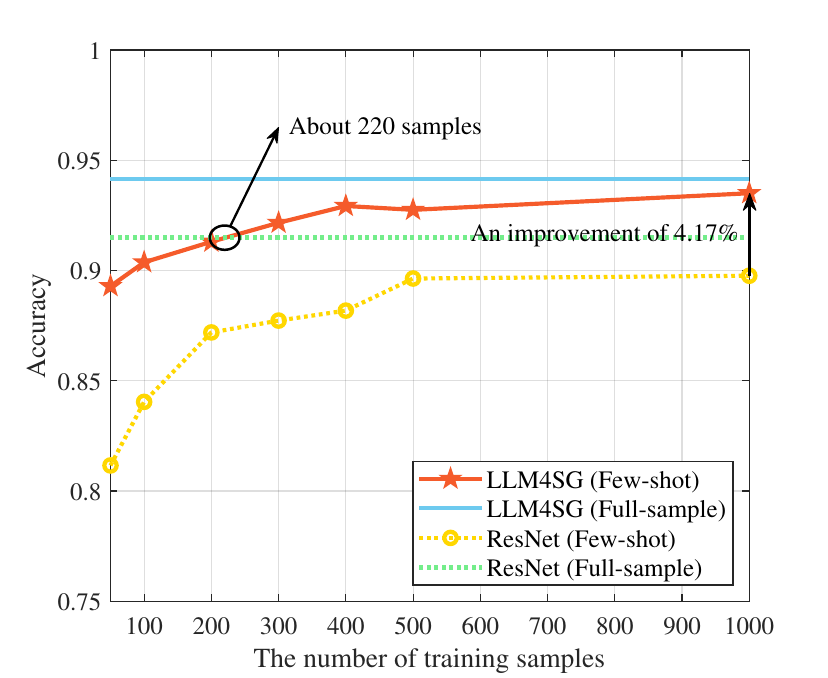}
        \centerline{(c)}
    \end{minipage}%
    \begin{minipage}[t]{0.5\linewidth}
        \centering
        \includegraphics[width=\textwidth]{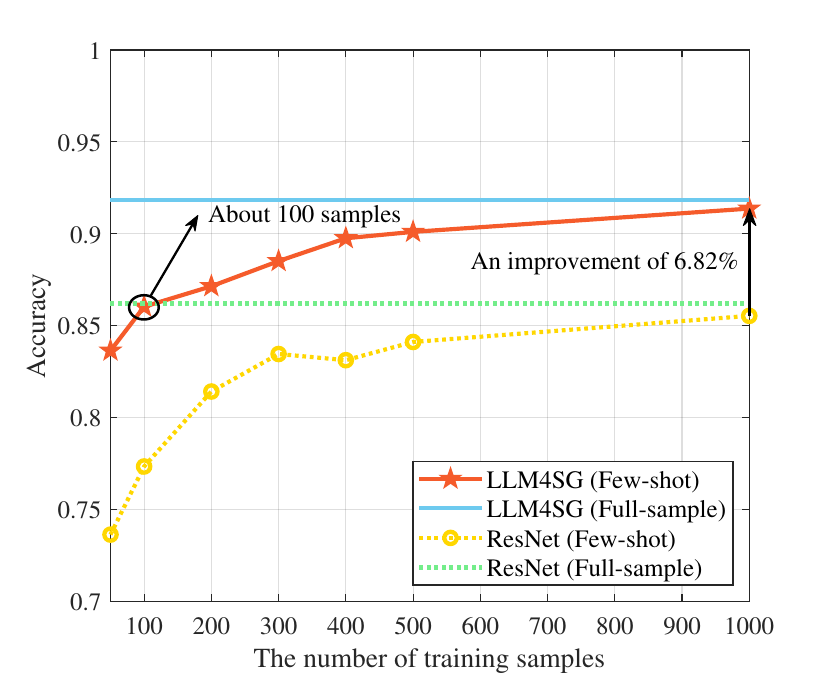}
        \centerline{(d)}
    \end{minipage}
    \caption{Generalization performance in scenario transfer. (a) Location evaluation metric from urban crossroad to urban wide lane. (b) Number evaluation metric from urban crossroad to urban wide lane. (c) Location evaluation metric from urban wide lane to urban crossroad. (d) Number evaluation metric from urban wide lane to urban crossroad.}
    \label{shi2chao}
\end{figure}

The performance of the LLM4SG model is evaluated in scenario transfer learning between three distinct environments, namely urban crossroad, suburban forking road, and urban wide lane~\cite{synthsom}. 
The model is first trained on one scenario and then fine-tuned with a limited number of samples from another scenario. 
Unlike the frequency band transfer, where the focus was on channel characteristics, this transfer challenges the model’s ability to adapt to entirely different spatial and environmental features. 
The performance of LLM4SG is compared to that of the baseline across different few-shot sample numbers, as shown in Fig.~\ref{shi2fen} and Fig.~\ref{shi2chao}.
Simulation results show that LLM4SG more effectively captures generalizable features across the aforementioned scenarios and demonstrates greater adaptability, especially when data is limited. 
With fewer than 420 samples (less than 2\% of the full dataset) for few-shot training, LLM4SG can already match the baseline’s performance under full-sample training.
This highlights the model’s ability to generalize across environments with varying road layouts and traffic patterns.
The complexity of the scenarios decreases from wide lane to crossroad and then to forking road. 
In this order, transferring from a more complex to a simpler scenario requires fewer training samples for LLM4SG to outperform ResNet trained with full samples. 
This is because complex scenarios include more varied objects, occlusions, signal reflections, and dynamic changes, which help the model learn more useful and general features.
In a word, this capability enables rapid synthesis of scenario-diverse scatterer datasets, reducing data collection cost and supporting 6G AI-native deployments.

\subsubsection{Performance Comparison in Vehicular Traffic Density Transfer}

\begin{figure}[t]
    \begin{minipage}[t]{0.5\linewidth}
        \centering
        \includegraphics[width=\textwidth]{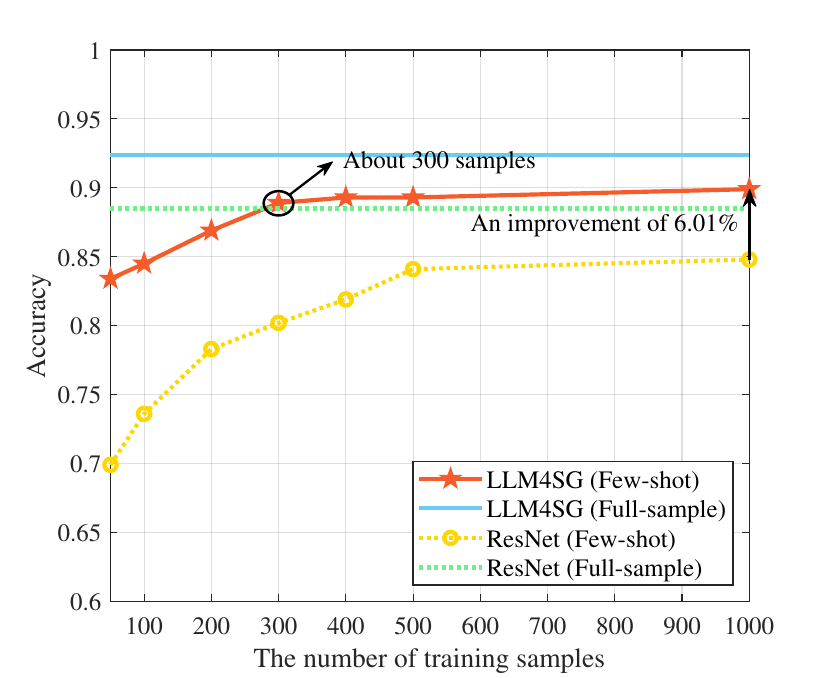}
        \centerline{(a)}
    \end{minipage}%
    \begin{minipage}[t]{0.5\linewidth}
        \centering
        \includegraphics[width=\textwidth]{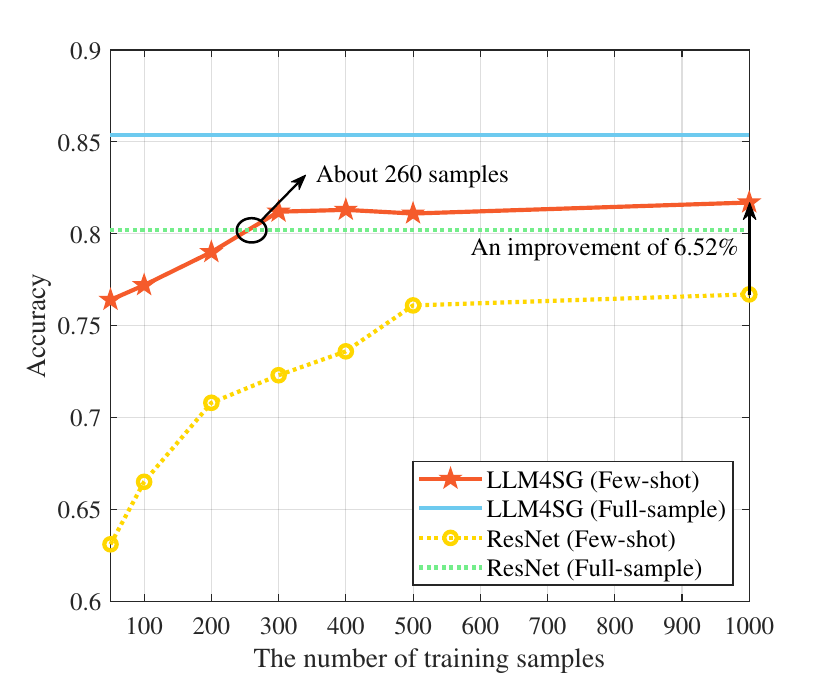}
        \centerline{(b)}
    \end{minipage}
    \hfill
    \begin{minipage}[t]{0.5\linewidth}
        \centering
        \includegraphics[width=\textwidth]{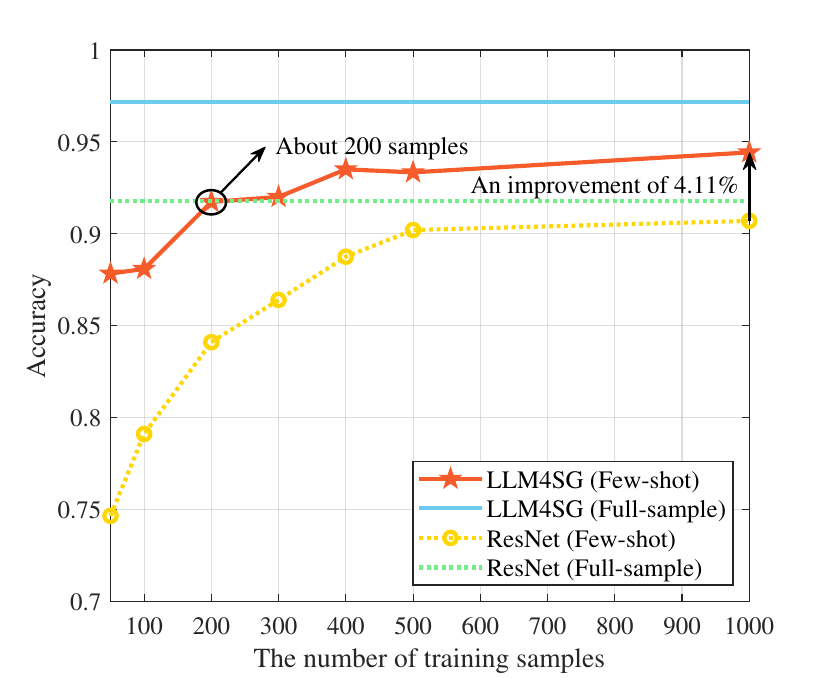}
        \centerline{(c)}
    \end{minipage}%
    \begin{minipage}[t]{0.5\linewidth}
        \centering
        \includegraphics[width=\textwidth]{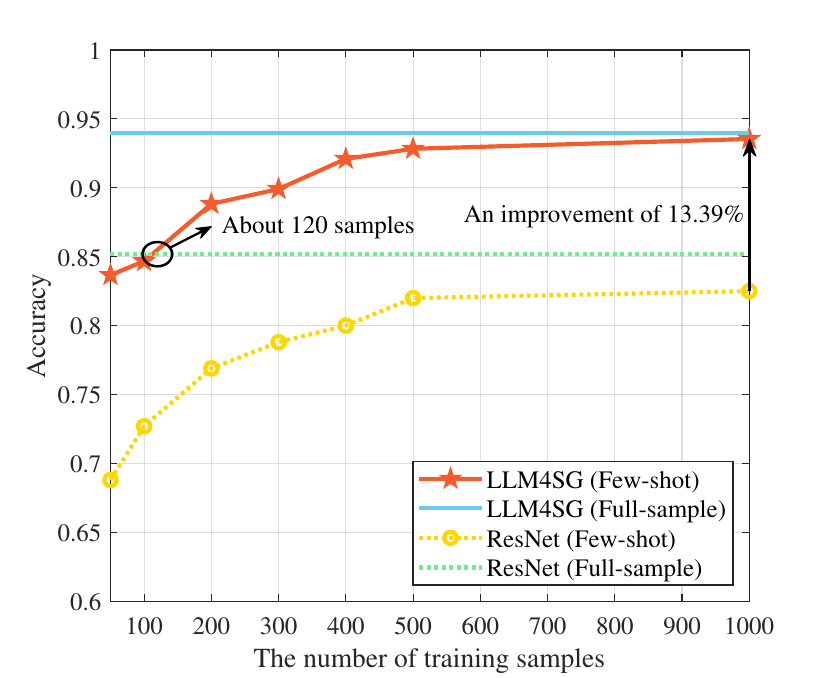}
        \centerline{(d)}
    \end{minipage}
    \caption{Generalization performance in VTD transfer in urban crossroad. (a) Location evaluation metric from low VTDs to high VTDs. (b) Number evaluation metric from low VTDs to high VTDs. (c) Location evaluation metric from high VTDs to low VTDs. (d) Number evaluation metric from high VTDs to low VTDs.}
    \label{low2high}
\end{figure}

The performance of the LLM4SG is evaluated on VTD transfer between high and low VTD conditions. 
The analysis evaluates the adaptability of LLM4SG to changing traffic conditions by training on one VTD and testing on another. 
The performance of LLM4SG is compared to that of the baseline across different few-shot sample numbers, as shown in Fig. \ref{low2high}.
Simulation results show that performance on the target VTD consistently improves with more transfer samples. 
Notably, LLM4SG achieves better adaptation across VTD conditions than the baseline, particularly with minimal few-shot data.
With fewer than 300 samples (less than 2\% of the full dataset) for few-shot training, LLM4SG can already match the baseline’s performance under full-sample training.
Furthermore, generalization from high VTDs to low VTDs requires fewer training samples compared to the reverse direction. 
This is because high VTD scenarios are inherently more complex, with richer multipath propagation and greater mobility, allowing the model to learn more comprehensive and transferable features during training.
To sum up, LLM4SG can generate VTD-diverse scatterer datasets with minimal few-shot data, providing the high-quality data substrate needed to train and deploy 6G AI-native systems.

\subsection{Ablation Experiments}
To assess the effectiveness of specific components, ablation experiments are performed by removing each key module, i.e., patching, positional embedding, and the LLM, as detailed in Table \ref{Ablation}.
\begin{table*}[t]
\renewcommand\arraystretch{1}
\setlength{\tabcolsep}{13.5pt}
\centering
\caption{Results of Ablation Experiments}
\label{Ablation}
\begin{tabular}{ccccccccccccc}
\hline
\multirow{2}{*}{Model} & \multicolumn{2}{c}{\begin{tabular}[c]{@{}c@{}}Crossroad\_28GHz\\ \_HighVTD\end{tabular}} & \multicolumn{2}{c}{\begin{tabular}[c]{@{}c@{}}Crossroad\_28GHz\\ \_LowVTD\end{tabular}} & \multicolumn{2}{c}{\begin{tabular}[c]{@{}c@{}}Crossroad\_sub6GHz\\ \_HighVTD\end{tabular}} & \multicolumn{2}{c}{\begin{tabular}[c]{@{}c@{}}Crossroad\_sub6GHz\\ \_LowVTD\end{tabular}} \\ \cline{2-9} 
& $P_{\rm pos}$          & $P_{\rm num}$                & $P_{\rm pos}$          & $P_{\rm num}$               & $P_{\rm pos}$           & $P_{\rm num}$              & $P_{\rm pos}$           & $P_{\rm num}$             \\ \hline
LLM4SG   & \textbf{94.1\%}& \textbf{91.8\%}&\textbf{96.3\%}& \textbf{94.5\%}&\textbf{92.3\%} &\textbf{85.3\%}  & \textbf{97.1\%} & \textbf{94.0\%}\\ \hline
w/o patching   & 91.6\%& 86.0\%&91.8\%& 87.2\%&88.2\% &79.8\%  & 92.1\% & 88.7\%\\ \hline
w/o positional embedding   & 68.5\%& 63.3\%&69.1\%& 65.4\%&62.7\% &58.3\%  & 65.4\% & 68.8\%\\ \hline
w/o LLM   & 83.4\%& 79.5\%&83.8\%& 81.5\%&79.3\% &74.9\%  & 82.5\% & 79.6\%\\ \hline
\end{tabular}
\end{table*}
The NMSE performance for the test set is recorded. 
For the LLM4SG without patching, the data was input directly without segmentation. 
For the LLM4SG without positional embedding, all position-specific information was excluded from the input data. 
For the LLM4SG without the LLM module, the pre-trained LLM was removed, with other network components left unchanged. 
The results show that removing any of these three components led to a decline in performance, underscoring the importance of each for achieving high predictive accuracy.

\subsection{Network Storage and Inference Cost}
The parameters of models and training and inference time cost by models are closely related to the storage and computational overhead.
The number of parameters and training and inference time of LLM4SG and the baseline are shown in Table \ref{cost}, where training and inference samples are taken from urban crossroad.
\begin{table}[t]
\caption{Network Parameters and Training and Inference Time Per Snapshot}
\label{cost}
\begin{tabular}{cccc}
\hline
       & Parameters (M) & Training time (ms) & Inference time (ms) \\ \hline
LLM4SG &   5.46/86.19  &   19.41    &    7.96    \\ \hline
ResNet &  23.53/23.53  &  22.8   &    5.67  \\ \hline
\end{tabular}
\end{table}
It can be observed that LLM4SG, with LoRA fine-tuning applied to only a small subset of parameters, incurs significantly lower training overhead compared to the baseline.
Meanwhile, LLM4SG has a comparable inference time with the baseline.

\subsection{Application Case}

\begin{figure}[t]
\centering
\includegraphics[width=3in]{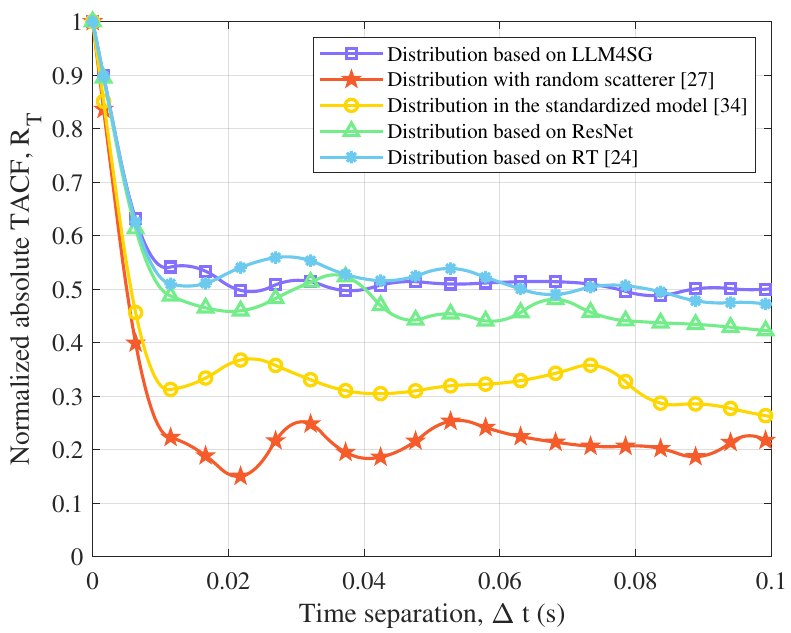}
\caption{Comparison of simulated TACFs and RT-based TACFs.}
\label{TACF_LLM}
\end{figure}

Conventional channel modeling can be classified into geometry-based deterministic modeling (GBDM), non-geometry stochastic modeling (NGSM), and geometry-based stochastic modeling (GBSM)~\cite{bai}.
The GBDM-based channel model is constructed in a deterministic manner for a specific scenario and leads to high complexity.
To reduce the complexity, the NGSM-based channel model determines channel parameters in a stochastic manner and the GBSM-based channel model utilizes the predefined stochastic distribution of scatterers/clusters in the propagation environment.
However, the accuracy of stochastic models is limited due to the random generation and determination of channel parameters.
LLM4SG provides a novel method to generate scatterers accurately.
We can exploit the predicted scatterer grid map $\hat{\boldsymbol{\phi_g}}(t)$ to generate scatterers instead of the predefined stochastic distribution.
Based on the channel model proposed in~\cite{huang2024lidar} and scatterers generated through different methods, the time auto-correlation function (TACF) generated based on the random statistical distribution of scatterers and TACF based on scatterers from LLM4SG are derived.
Then, the RT-based TACF~\cite{InSite} and TACF obtained from standardized models~\cite{generation2019technical} are compared with the aforementioned TACF, as shown in Fig. \ref{TACF_LLM}.
It can be seen that the TACF based on scatterers from LLM4SG that captures time consistency can achieve a close fit with the RT-based TACF. 


\section{Conclusions}
Based on a new synthetic intelligent sensing-communication dataset, i.e., SynthSoM-V2V, a novel method, named LLM4SG, has been developed to generate scatterers from LiDAR point clouds.
In LLM4SG, LoRA fine-tuning has been applied to carefully selected modules in a task guided manner, enabling intelligent sensing-communication integration while leveraging the LLM’s powerful cross-modal representation capabilities.
Simulation results have indicated that the proposed LLM4SG has delivered outstanding performance across full-sample and generalization testing.
The developed LLM4SG has achieved over 92\% accuracy in scatterer position generation and over 85\% accuracy in scatterer quantity generation.
Moreover, the results have demonstrated the strong few-shot and generalization capabilities of LLM4SG, with the model consistently surpassing conventional deep learning models in knowledge transfer across different VTDs, frequency bands, and V2V scenarios, showing an average improvement of over 7.5\%.
The advanced architectures of LLMs provide deep contextual understanding and strong adaptability, enabling them to handle varying conditions with minimal retraining. 
We have leveraged sensing data that faithfully captures the physical environment, which can be acquired at relatively low cost and thereby enable scalable data generation.
This flexibility is crucial for meeting the massive and diverse data requirements of AI-native 6G communication systems, ensuring reliable generation in high-mobility real-world V2V scenarios.
In LLM4SG, dedicated modules are designed to strategically leverage these strengths for accurate scatterer generation.


\end{document}